%% file: SuGOHIV_acc.tex
\documentclass[fleqn,usenatbib]{mnras}
\usepackage{newtxtext,newtxmath}
\usepackage[T1]{fontenc}
\usepackage{graphicx}	
\usepackage{amsmath}	
\usepackage{amssymb}	
\usepackage{float}
\usepackage[style=base]{caption}
\usepackage{multirow}
\usepackage{hyperref}
\usepackage{natbib}
\usepackage[usenames]{color}
\graphicspath{{figs/}}
\usepackage[dvipsnames]{xcolor}
\usepackage{array}
\usepackage{footmisc}
\usepackage{balance}
\newcommand{\K}[1]{{#1}$^{\mathbb{K}}$}
\newcommand{\C}[1]{{#1}$^{\mathbb{C}}$}
\newcommand{\D}[1]{{#1}$^{\mathbb{D}}$}

\newcolumntype{C}[1]{>{\centering\arraybackslash}m{#1}}
\newcolumntype{L}[1]{>{\arraybackslash}m{#1}}

\input{macros.tex}

\title[SuGOHI V: Group-to-cluster scale lens search]{Survey~of~Gravitationally-lensed~Objects~in~HSC~Imaging~(SuGOHI). V. Group-to-cluster scale lens search from the HSC-SSP Survey}
\author[A. T. Jaelani et al.]{Anton T. Jaelani$^{1,2,3}$\thanks{E-mail: \href{anton@phys.kindai.ac.jp}{ anton@phys.kindai.ac.jp}},
Anupreeta More$^{4,5}$, 
Masamune Oguri$^{4,6,7}$,
Alessandro Sonnenfeld$^{4,8}$,\newauthor
Sherry H. Suyu$^{9,10,11}$,
Cristian E. Rusu$^{12}$,
Kenneth C. Wong$^{4}$,
James H. H. Chan$^{13}$, \newauthor
Issha Kayo$^{14}$, 
Chien-Hsiu Lee$^{15}$,
Dani C. -Y. Chao$^{9,10}$, 
Jean Coupon$^{16}$, 
Kaiki T. Inoue$^{1}$, \newauthor
and Toshifumi Futamase$^{17}$
\\ 
$^{1}$Department of Physics, Kindai University, 3-4-1 Kowakae, Higashi-Osaka, Osaka 577-8502, Japan\\
$^{2}$Astronomical Institute, Tohoku University, 6-3 Aramaki, Aoba-ku, Sendai 980-8578, Japan\\
$^{3}$Astronomy Study Program and Bosscha Observatory, FMIPA, Institut Teknologi Bandung, Jl. Ganesha 10, Bandung 40132, Indonesia\\
$^{4}$Kavli Institute for the Physics and Mathematics of the Universe	(IPMU), 5-1-5 Kashiwanoha, Kashiwa-shi, Chiba 277-8583, Japan\\
$^{5}$The Inter-University Centre for Astronomy and Astrophysics (IUCAA), Post Bag 4, Ganeshkhind, Pune 411007, India\\
$^{6}$Department of Physics, The University of Tokyo, 7-3-1 Hongo, Bunkyo-ku, Tokyo 113-0033, Japan\\
$^{7}$Research centre for the Early Universe (RESCEU), The University of Tokyo, 7-3-1 Hongo, Bunkyo-ku, Tokyo 113-0033, Japan\\
$^{8}$Leiden Observatory, Leiden University, Niels Bohrweg 2, 2333 CA Leiden, the Netherlands\\
$^{9}$Max-Planck-Institut f\"{u}r Astrophysik, Karl-Schwarzschild-Stra{\ss}e 1, 85748 Garching, Germany\\
$^{10}$Physik-Department, Technische Universit\"{a}t M\"{u}nchen, James-Franck-Stra{\ss}e 1, 85748 Garching, Germany\\
$^{11}$Academia Sinica Institute of Astronomy and Astrophysics (ASIAA), 11F of ASMAB, No. 1, Section 4, Roosevelt Road, Taipei 10617, Taiwan\\
$^{12}$Subaru Telescope, National Astronomical Observatory of Japan, 2-21-1 Osawa, Mitaka, Tokyo 181-0015, Japan\\
$^{13}$Laboratory of Astrophysique, \'Ecole Polytechnique F\'ed\'erale de Lausanne (EPFL), Observatoire de Sauverny, 1290 Versoix, Switzerland\\
$^{14}$Department of Liberal Arts, Tokyo University of Technology 5-23-22 Nishikamata, Ota-ku, Tokyo 144-8650, Japan\\
$^{15}$National Optical Astronomy Observatory 950 N Cherry Avenue, Tucson, AZ 85719, USA\\
$^{16}$Department of Astronomy, University of Geneva, ch. d'\'Ecogia 16, 1290 Versoix, Switzerland\\
$^{17}$Department of Astrophysics and Atmospheric Sciences, Kyoto Sangyo University, Kyoto, Kyoto 603-8555, Japan}
\date{Accepted XXX. Received YYY; in original form ZZZ}

\pubyear{2020}

\begin{document}
\label{firstpage}
\pagerange{\pageref{firstpage}--\pageref{lastpage}}
\maketitle

\begin{abstract}
We report the largest sample of candidate strong gravitational lenses belonging to the Survey of Gravitationally-lensed Objects in HSC Imaging for group-to-cluster scale (SuGOHI-c) systems. These candidates are compiled from the S18A data release of the Hyper Suprime-Cam Subaru Strategic Program (HSC-SSP) Survey. We visually inspect $\sim39,500$ galaxy clusters, selected from several catalogs, overlapping with the Wide, Deep, and UltraDeep fields, spanning the cluster redshift range $0.05<\zcl<1.38$. We discover 641 candidate lens systems, of which 536 are new. From the full sample, 47 are almost certainly bonafide lenses, 181 of them are highly probable lenses and 413 are possible lens systems. Additionally, we present 131 lens candidates at galaxy-scale serendipitously discovered during the inspection. We obtained spectroscopic follow-up of 10 candidates using the X-shooter. With this follow-up, we confirm 8 systems as strong gravitational lenses. Of the remaining two, one of the sources is too faint to detect any emission, and the other has a tentative redshift close to the lens redshift, but additional arcs in this system are yet to be observed spectroscopically. Since the HSC-SSP is an ongoing survey, we expect to find $\sim600$ definite or probable lenses using this procedure and even more if combined with other lens finding methods.
\end{abstract}


\setcounter{footnote}{1}

\section{Introduction} \label{sec:intro}
The standard model of cosmology suggests that the Universe is dominated by dark matter and dark energy. Strong gravitational lensing is a phenomenon where multiply-lensed images of distant sources can be seen due to deflection by the gravity of intervening massive objects such as galaxies and galaxy clusters. Gravitational lensing has been shown to be a promising technique to probe these dark components. Lensing has been used to study distant galaxies with extreme magnification \cite[e.g.,][]{Swinbank+09,Zitrin+09,Richard+11}, infer substructure in the lensing halos \cite[e.g.,][]{More+09,Vegetti+10,Vegetti+10b,Hezaveh+16}, constrain the Hubble constant \cite[e.g.,][]{Suyu+10,Bonvin+17,Wong+19} and place constraints on the slope of the inner density profile of the lensing halos \cite[e.g.,][]{Koopmans+03,Koopmans+06,More+08,Barnabe+09,Koopmans+09}.

This has motivated dedicated efforts to search for gravitational lenses in large astronomical surveys e.g., the Hyper Suprime-Cam Subaru Strategic Program (HSC-SSP) Survey \cite[][]{Aihara+18}, DESI Legacy Imaging Surveys \cite[][]{Dey+19}, Kilo Degree Survey \cite[KiDS,][]{deJong+15}, and Dark Energy Survey \cite[DES,][]{DES+16}. Specifically, large imaging and spectroscopic surveys have allowed inferences of statistical properties of lenses such as constraints on the stellar initial mass function \cite[e.g.,][]{Treu+10,Ferreras+10,Sonnenfeld+12,Sonnenfeld+19}, estimation of the fraction of dark matter in galaxy-scale halos \cite[e.g.,][]{Gavazzi+07,Grillo+10,Faure+11,Ruff+11,More+11}, and even constraints on cosmology \cite[e.g.,][]{Gladders+03,Oguri+12}.

As mentioned above, most of the surveys have primarily focused on studying galaxy-scale or cluster-scale structures. As a result, the matter distribution in galaxies and galaxy clusters is relatively well-studied via both strong and weak lensing. A further improvement in our understanding has come from the use of complementary methods to lensing such as stellar kinematics, satellite kinematics, and X-ray scaling relations. In contrast, there has not been much progress (in the last decade) on mass distributions of galaxy groups in the mass range of 10$^{12} -10^{14} \Msun$, intermediate to galaxies and galaxy clusters. Using X-ray samples to study the intra-group medium at low redshifts \citep{Helsdon+00}, mass-to-light ratios of groups from the Canadian Network for Observational Cosmology 2 (CNOC 2) sample \cite[e.g.,][]{Parker+05}, the faint end of the luminosity function of nearby compact groups \cite[e.g.,][]{Krusch+06}, the concentration-mass ($c-M$) relation of groups \cite[e.g.,][]{Mandelbaum+08,Newman+15}, colours and star formation of galaxy groups \cite[e.g.,][]{Balogh+09,Balogh+11}, scaling relations of X-ray selected groups \citep{Rines+10}, and baryon fractions from the Two Micron All Sky Survey (2MASS) \citep{Dai+10} are some examples of investigations of galaxy groups.

We still do not have a detailed understanding of the matter distribution, formation, and evolution of galaxy groups. Being one of the important components in the hierarchical assembly of structures in the Universe, galaxy groups are much more massive than galaxy-scale halos and are concentrated enough to act as lenses. Furthermore, since galaxy groups are quite abundant compared to massive structures like galaxy clusters, the probability of finding group-scale lenses is also large. Hence, strong lensing can be successfully used to study group-scale halos \citep{Limousin+09,More+12,Foex+13,Foex+14,Verdugo+14,Newman+15}.

In this work, we conduct a systematic search of group- and cluster-scale lenses as part of the Survey of Gravitationally-lensed Objects in HSC Imaging (SuGOHI) and also present the results of spectroscopic follow-up of a sub-sample of these systems. The galaxy-scale lens sample (SuGOHI-g) is presented in \citet{Sonnenfeld+18,Sonnenfeld+19} and \citet{Wong+18}, and the results of a search for lensed quasars (SuGOHI-q) are reported in \citet{Chan+19}. This paper is organised as follows. In \sref{sec:data}, we describe the HSC-SSP imaging data used in our search. In \sref{sec:method}, we describe the procedure for finding new strong gravitational lens systems. We present our newly discovered lens candidates in \sref{sec:results}. In \sref{sec:follow}, we describe our spectroscopic follow-up observation. We present our summary and conclusion in \sref{sec:summary}. 

\begin{figure*}
\includegraphics[width=0.90\textwidth]{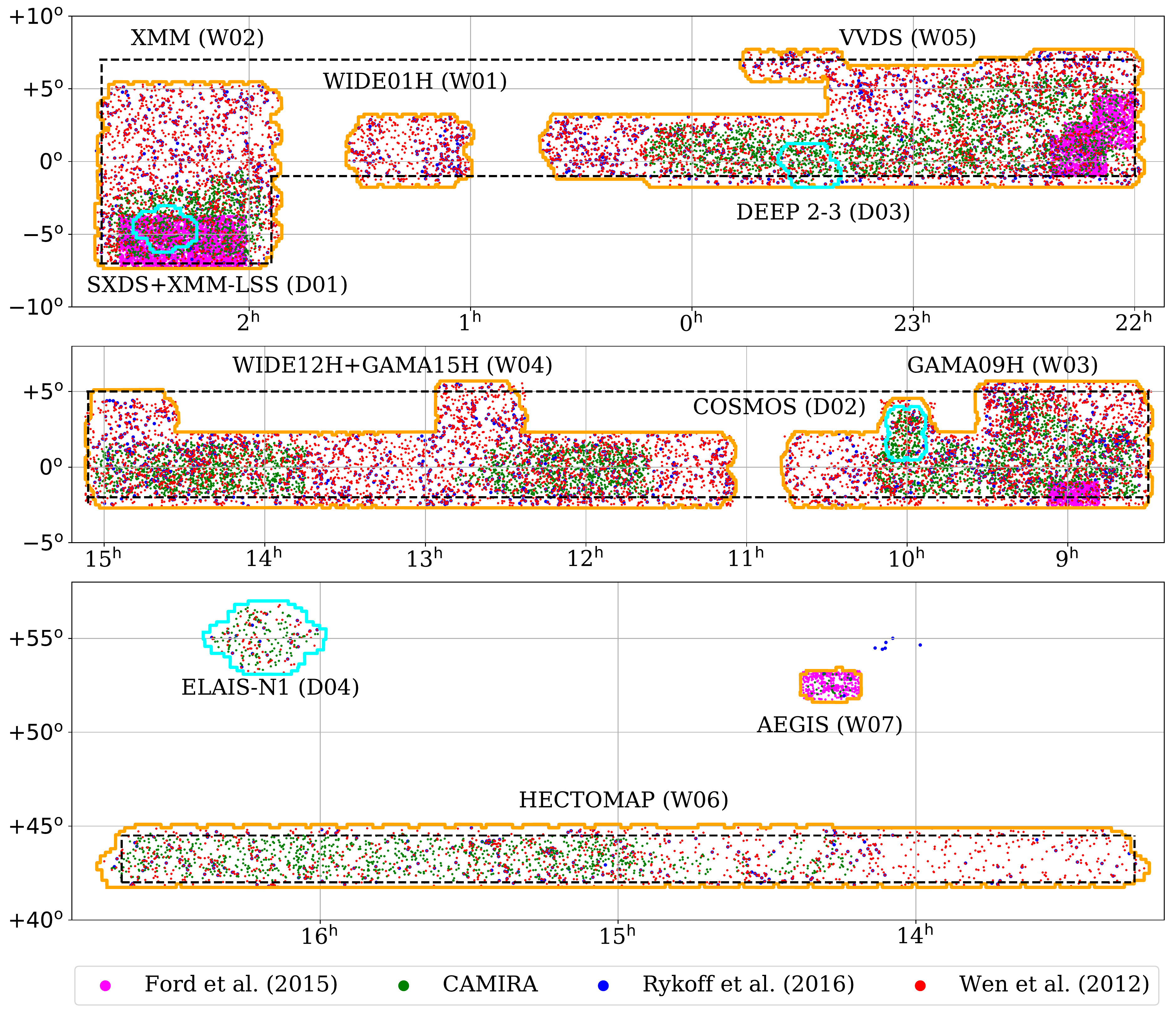}
\caption{The Hyper Suprime-Cam Subaru Strategic Program (HSC-SSP) observational footprint shown in equatorial coordinates. The orange and cyan boxes indicate the Wide and Deep+UltraDeep fields for the S18A data release (internal), respectively. The dashed black boxes indicate the approximate boundaries of the three disjoint regions that will make up the final Wide survey. The overlapping cluster catalogs are shown by different point colours.}\label{fig:Figure1}
\end{figure*}

\begin{table}
\caption{Clusters found in the HSC-SSP S18A footprint from different algorithms, some of which are external to the HSC Survey collaboration.}
\label{tab:Table1}
\input{tabs/Table1.tex}
\end{table}

\section{The Data} \label{sec:data}
The Subaru Strategic Program (SSP) survey is carried out with the Hyper Suprime-Cam \cite[HSC,][]{Miyazaki+18,Komiyama+18, Kawanomoto+18, Furusawa+18,Huang+18,Coupon+18}, a $1.7\deg^2$ field-of-view optical camera recently installed on the 8.2-m \textit{Subaru telescope}. The HSC-SSP Survey has three fields; the Wide field is expected to cover a 1,400 deg$^2$ area in five bands ($g$, $r$, $i$, $z$, and $y$) to an $i$-band depth of 26.2 by its completion, while the Deep+UltraDeep fields are expected to cover smaller areas of about 27~deg$^2$ and 3.5~deg$^2$, respectively (see \citet{Aihara+18} for more details about the survey). We use the photometric data from the S18A data release, which covers 1,114 deg$^2$ (out of which 305 deg$^2$ is full depth) in Wide and 31 deg$^2$ in Deep+UltraDeep, at least in one filter and one exposure \citep{Aihara+19}. The data are processed with the reduction pipeline \textsc{hscPipe v6.7} \citep{Bosch+18}, a version of the Large Synoptic Survey Telescope stack \cite[][]{Axelrod+10,Juric+17,Ivezic+08,Ivezic+19}. The median seeing of S18A data is 0.61 arcsec in the $i$-band, 0.85 arcsec in the $g$-band and the pixel scale of HSC is 0.168 arcsec.

The redshifts used in this work are obtained from the photometric redshift catalog of the HSC-SSP Survey, determined using the Direct Empirical Photometric code \citep[\texttt{DEmP},][]{Hsieh+14}. The HSC-SSP photometric redshifts are most accurate at $0.2\lesssim z_{\rm phot}\lesssim1.5$. The point estimates of the photometric redshift are accurate to better than 1\% in term of $\langle \Delta z/(1+z)\rangle$ with a scatter of $\approx0.04$ and an outlier rate of $\approx8\%$ for galaxies with $i<24$ mag. A more detailed description of \texttt{DEmP}'s application to the HSC-SSP data is presented in \cite{Tanaka+18}. Since the HSC-SSP Survey footprint has some overlap with that of the Sloan Digital Sky Survey (SDSS), we also extracted spectroscopic redshifts, whenever available, from the SDSS Data Release 15 \citep{Aguado+19} catalogs.


\section{Lens Candidate Selection} \label{sec:method}
In this section, we describe how the cluster catalogs are selected for visual inspection and our criteria to grade the lens candidates. Visual inspection of cluster catalogs is still a practically useful method for lens finding at group/cluster scales because it allows for selection of candidates with rich diversity and complexity while maintaining high purity. We used the \textsc{hscMap (Sky Explorer)\footnote[1]{\label{note1}\href{https://hsc-release.mtk.nao.ac.jp/hscMap-pdr2/app/}{https://hsc-release.mtk.nao.ac.jp/hscMap-pdr2/app/}, similar tool but for public data release 2.}}, an online tool, to visually inspect colour images of the clusters (see \sref{sec:parent}). The lens search relied on morphology and colour information to visually analyse properties of the lensed images and the galaxies in the cluster to assess the plausibility of lensing. Inspectors could control the spatial scales, contrast, and brightness, and could choose different combinations of the HSC filters.

\subsection{Parent Catalogs}\label{sec:parent}
We used galaxy cluster catalogs within the footprint covering the HSC-SSP S18A imaging. The on-sky distribution of these clusters, along with survey footprints, is shown in \fref{fig:Figure1}. We also give the number of clusters detected from each of the four catalogs in \tref{tab:Table1}.

\subsubsection{Clusters from the HSC-SSP Survey}

\begin{figure*}
\includegraphics[width=\textwidth]{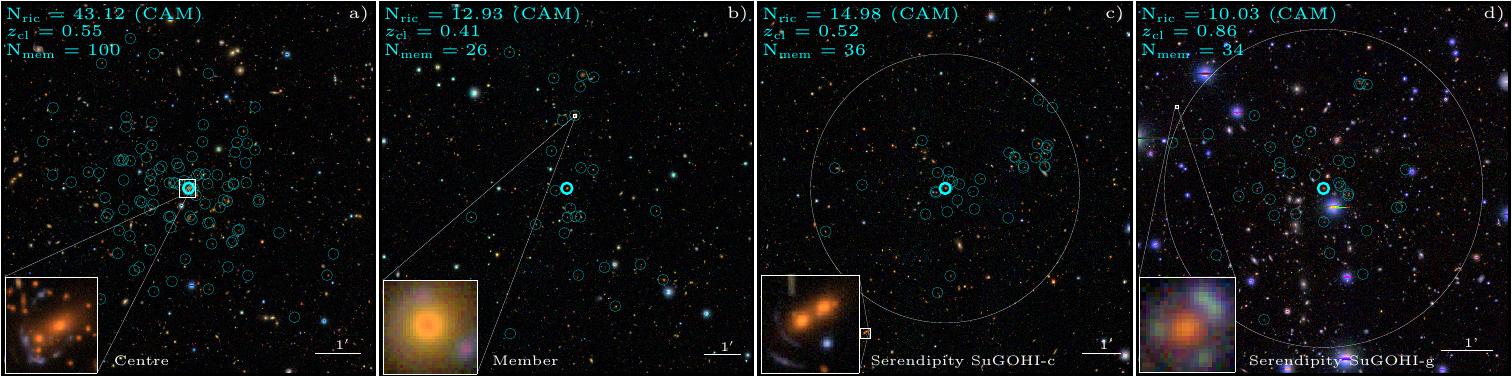}
\caption{Types of lens candidates depending on their location with respect to galaxy clusters. Lens candidates where the BCG acts as a lens (\textbf{panel a}, e.g., HSC J1441$-$0053) or a member galaxy (see more description in \sref{sec:results}) acts as a lens (\textbf{panel b}, e.g., HSC J1234$-$0009) are classified as SuGOHI-g, otherwise as serendipitous discoveries (\textbf{panel c and d}, e.g., HSC J1414$-$0136 and HSC J0904$+$0102, for SuGOHI-c and -g, respectively). The cyan circles indicate member galaxies of a cluster and the brightest cluster galaxy (BCG) is indicated by a thicker circle. The white circle indicates the region which encloses member galaxies with the BCG at the centre. The richness $N_{\rm ric}$, cluster redshift $\zcl$, and number of galaxy member $N_{\rm mem}$, are shown at the top left.}
\label{fig:Figure2}
\end{figure*}

Our primary cluster catalog is called \textsc{camira} which is produced by running the cluster-finding algorithm \citep{Oguri+14} on the internal HSC-SSP data release S18A\footnote[2]{Note that the currently published \textsc{camira} catalog makes use of the data release S16A only and it can be obtained from \citet{Oguri+18}. However, this is a subsample of the catalog used in our study.} \citep{Aihara+19}, covering roughly 465 deg$^2$ and 28 deg$^2$, in all five filters, for Wide and Deep+UltraDeep fields, respectively. The \textsc{camira} is validated through comparisons with existing spectroscopic and X-ray data as well as mock galaxy catalogs.

We obtain 14,992 clusters, comprising of 14,350 clusters from the Wide fields and 642 clusters from the Deep fields, with the richness limit $N_{\rm ric, \textsc{camira}}>10$ spanning a redshift range of $0.1<\zcl<1.38$. Richness in \textsc{camira} is defined to be the number of red member galaxies with stellar mass $M_{\rm star}\geq 10^{10.2}\Msun$ lying within a physical radius of $\approx1.4$ Mpc. The richness limit $N_{\rm ric, \textsc{camira}}= 10$ corresponds to $M_{200}\approx 7\times 10^{13} \Msun$, where $M_{200}$ is the cluster mass within $r_{200}$, by extrapolating the richness-mass relation of \textsc{camira} from \citet{Murata+19}. $r_{200}$ is the radius within which the mean density of a cluster is 200 times the critical density of the Universe. The cluster sample is shown by green points in \fref{fig:Figure1}. The \textsc{camira} cluster catalog is still being constructed from the HSC-SSP Survey data, since the HSC survey is an ongoing and several patches are likely to have incomplete imaging data, e.g., has $\leq 4$ filters. As a result, we decided to make use of other public catalogs constructed from previous surveys which have overlapped with the HSC-SSP Survey footprint, e.g., the Sloan Digital Sky Survey \citep[SDSS,][]{York+00} and the Canada-France Hawaii Telescope Lensing Survey \citep[CFHTLenS,][]{Heymans+12}. Also, including more cluster catalogs maximises the chance of finding more group-to-cluster scale lenses.

\subsubsection{Clusters from Data Release 8 of SDSS-III Data}
The HSC-SSP Survey footprint has almost complete overlap with SDSS footprint. We thus have two extensive cluster catalogs, \citet{Wen+12} and \citet{Rykoff+16}, that can be used. Both catalogs are derived from the galaxy data of 14,000 deg$^2$ of SDSS-III \citep{Eisenstein+11}. \citet{Wen+12} identified 132,684 clusters (12,000 of them overlap with the HSC-SSP S18A footprint, see \tref{tab:Table1} and red points in \fref{fig:Figure1}) in the redshift range of $0.05\leq \zcl<0.8$. The clusters are selected if their richness $N_{\rm ric, Wen} \geq 12$ which corresponds to $M_{200}~\approx~0.6\times 10^{14} \Msun$ and a number of member galaxies candidates $N_{200}\geq 8$ within $r_{200}$. We also used clusters from the red-sequence Matched-filter Probabilistic Percolation \citep[\textsc{redMaPPer}, for the details see][]{Rykoff+14} cluster finding algorithm (version 6.3). This catalog has a total of 25,236 clusters (2,968 are overlapping with the HSC-SSP S18A footprint, see \tref{tab:Table1} and blue points in \fref{fig:Figure1}) in the redshift range $0.08\leq \zcl<0.55$ with $N_{\rm ric, \textsc{redMaPPer}}\geq 20$ which corresponds to $M_{200}\gtrsim 10^{14} \Msun$. For more detailed description of the catalog, see \cite{Rykoff+16}.

\subsubsection{CFHTLenS Data}
\citet{Ford+15} have a sample of 18,056 clusters (9,475 of them overlap with the HSC-SSP S18A footprint, see \tref{tab:Table1}) at redshifts $0.2\leq \zcl\leq 0.9$. The clusters have been detected using the 3D-Matched-Filter Galaxy Cluster Finder in the $\sim154$ deg$^2$ CFHTLenS survey \citep{Milkeraitis+10,Ford+14} with a significance $\geq 3.5$ and richness $N_{\rm ric, Ford}> 2$ which corresponds to $M_{200}~\approx~6 \times 10^{12} \Msun$. This field has a substantial overlap with the S18A data (see magenta points in \fref{fig:Figure1}).

\begin{table}
\caption{Lens candidate statistics. "\textsc{Cam}~F~R~W" represent the parent cluster catalogs as presented in \tref{tab:Table1}.}
\label{tab:Table2}
\input{tabs/Table2.tex}
\end{table}

\subsection{Ranking criteria}
We identified 39,435 clusters, located in the HSC-SSP S18A footprint, from the four catalogs combined. Inspectors use the online \textsc{hscMap (Sky Explorer)\footref{note1}} server to inspect colour images of the clusters \citep{Aihara+18b,Aihara+19}. In the first step, we divided the clusters into three redshift bins which were inspected by three inspectors per redshift bin. After we compiled the 1160 candidates from all redshift bins combined, nine inspectors independently assigned a rank from 0 to 3, according to the following criteria:

\begin{itemize}
\item 3: almost certainly a lens, 
\item 2: probably a lens, 
\item 1: possibly a lens, and
\item 0: not a lens.
\end{itemize}

We further refined the sample of candidates by applying the following scheme: 
\begin{enumerate}
\item[] A: $\rm{\langle Rank\rangle}>2.5$,
\item[] B: $1.5<\rm{\langle Rank\rangle}\leq2.5$, 
\item[] C: $0.5<\rm{\langle Rank\rangle}\leq1.5$, and
\item[] Not a lens: $\rm{\langle Rank\rangle}\leq0.5$
\end{enumerate}
where $\rm{\langle Rank\rangle}$ is the mean rank given by individual inspectors. The final sample, thus, consists of 772 candidates. Systems with highly discrepant ranks were discussed and regraded to mitigate such discrepancies.


\section{Results}\label{sec:results}

The Einstein radius, $\theta_{\rm Eins}$, is the best parameter to represent mass of the lens which can be approximated from the arc radius, R$_{\rm arc}\approx 2\theta_{\rm Eins}$ \citep{More+12}. Typically, lensing halos with $\theta_{\rm Eins}$ $\geq2$ arcsec are very massive lenses with significant contribution from the environment of the primary lensing galaxy \citep{Oguri+06,More+12}. Here, we calculated the arc radius by assuming a circle roughly covering the candidate arc centred on the Brightest Cluster Galaxy (BCG).

Next, we separated the graded systems into two groups: SuGOHI-c and additional lenses at galaxy-scale (SuGOHI-g, see \aref{app:sugohig}), which are shown by \tref{tab:Table3} and \tref{tab:TableA2}, respectively. The classification criteria for a candidate to be included in SuGOHI-c are the following: 
\begin{enumerate}
\item[1.] If the lensing is due to the brightest central galaxy (BCG) (see the panel a in \fref{fig:Figure2}), and
\item[2.] either the angular separation of the arc from the lens centre, the arc radius, R$_{\rm arc}\geq2$ arcsec (e.g., HSC J1557$+$4206),
\item[3.] or if the lensing is caused by more than one galaxy enclosed by a ring through the arc or multiple images (e.g., HSC J2228$+$0022).
\end{enumerate}

If none of the above are satisfied, candidates fall in the SuGOHI-g sample as being serendipitously discovered during the inspection (e.g., HSC J0904$+$0102 \citep{Jaelani+19}, which is at a similar position to HSC J1414$-$0136 with respect to the cluster but has R$_{\rm arc}<2$ arcsec; see the panel c of \fref{fig:Figure2}), and are reported in the Appendix.

For the first classification criteria, if a BCG is misclassified\footnote[3]{A BCG is considered to be misclassified if it is visually much brighter than its neighbours and/or the galaxy labelled as BCG by the cluster-finding algorithm.} as a member galaxy by the algorithm, only then do we accept it as a SuGOHI-c system. A member galaxy may be aided by the group potential, but then the arc separation also needs to be R$_{\rm arc}\geq2$ arcsec (see the middle panel in \fref{fig:Figure2}). Otherwise, this could still be a galaxy-scale lens. In some rare cases, an arc radius cannot be quantified because it is being deflected by multiple galaxies on either side (e.g., see HSC J0209$-$0448, grade C in the online material\footnote[4]{\href{http://www-utap.phys.s.u-tokyo.ac.jp/~oguri/sugohi/}{http://www-utap.phys.s.u-tokyo.ac.jp/$\sim$oguri/sugohi/}}). These are also included as SuGOHI-c.

\begin{figure}
\includegraphics[width=0.45\textwidth]{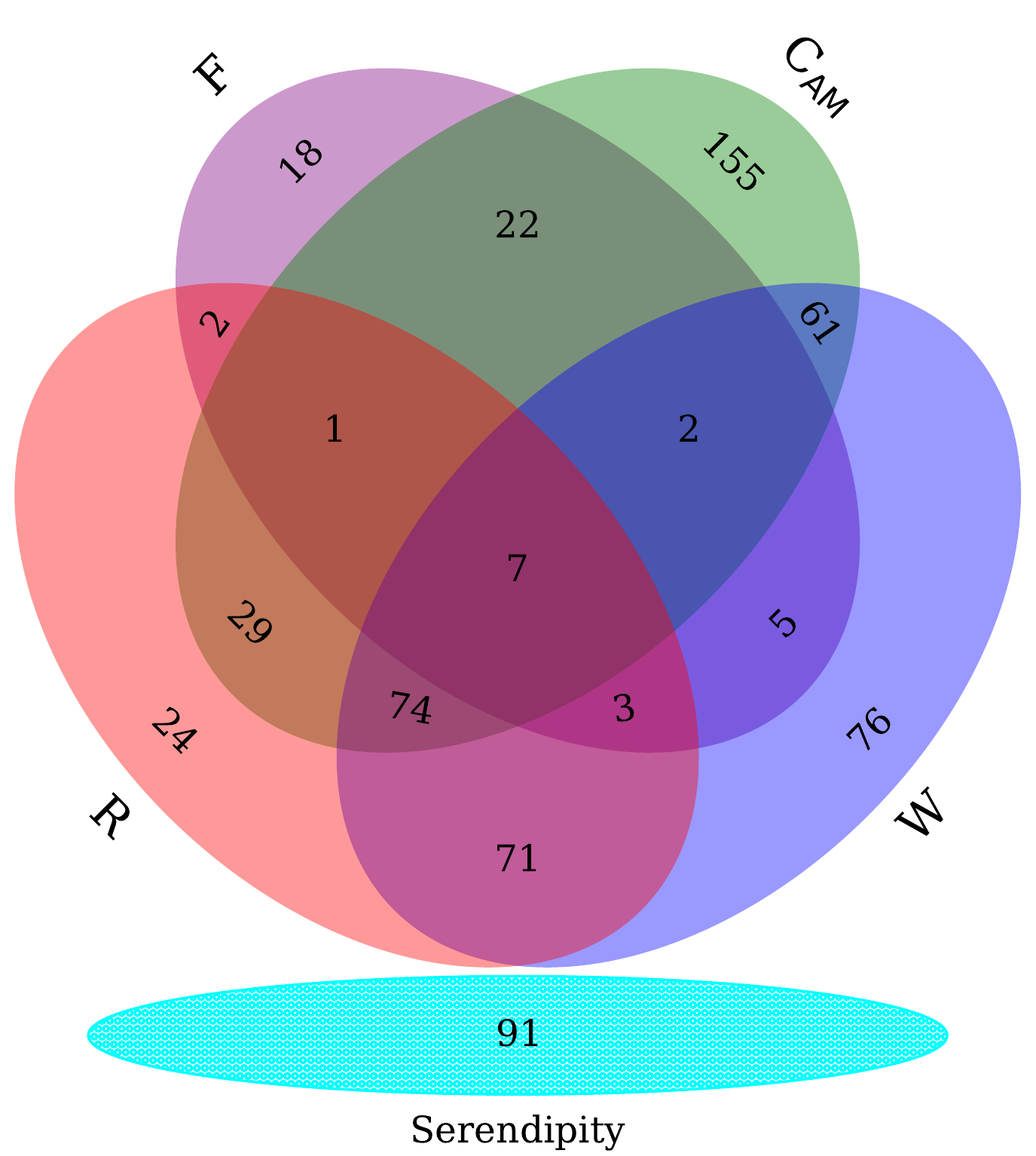}
\caption{Distribution of lens candidates according to the parent cluster catalogs. The letters represent the parent cluster catalogs as presented in \tref{tab:Table1}. The Venn diagram is divided into two panels: 550 lens systems of SuGOHI-c correspond to the parent cluster (upper) and 91 lens systems which are serendipiously (bottom) discovered during the inspection, and thus, are not listed as BCGs of the clusters or members with R$_{\rm arc}\geq2$ arcsec.}\label{fig:Figure3}
\end{figure}

A total of 641 systems (including 536 new lenses presented for the first time) are in the SuGOHI-c sample. These consist of 47 Grade A, 181 Grade B, and 413 Grade C systems, respectively. We found some candidate systems in more than one catalog that are shown in \fref{fig:Figure3}. The \textsc{camira} produces the largest number of candidate systems. We also found many candidates serendipitously which were missed by the parent cluster catalogs. Some of these lenses were also discovered independently by the citizen science project \citep[\textsc{Space Warps},][]{Marshall+16,More+16b} from the HSC-SSP Survey (Sonnenfeld et al., in prep). We present the lens candidate statistics in \tref{tab:Table2}. 

\begin{table*}
\caption{SuGOHI-c candidates, with grades A and B, selected by visual inspection of galaxy cluster catalogs. Redshifts are \texttt{DEmP} photometric and SDSS DR15 spectroscopic redshifts. PC indicates the parent cluster catalog through which a candidate was selected as in \tref{tab:Table1}. Systems with $\dagger$, $^{\rm X}$, $^{\mathbb{C}}$, $^{\mathbb{D}}$, and $^{\mathbb{K}}$ are the lens candidates which have red-coloured sources, X-shooter follow-up, spectroscopically confirmed, previously discovered, and known, respectively. Systems with references are previously known, whereas other objects with "..." are new. References: $^1$\citet{Diehl+17}, $^2$\citet{Huang+19}, $^3$\citet{Jacobs+19}, $^4$\citet{Petrillo+19}, $^5$\citet{Sonnenfeld+18}, $^6$\citet{Wong+18}, $^7$\citet{More+12}, $^8$\citet{More+16b}, $^9$\citet{Cabanac+07}, $^{10}$\citet{Stark+13b}, $^{11}$\citet{Tanaka+16}, $^{12}$\citet{Bolton+08}, $^{13}$\citet{Faure+08}, $^{14}$\citet{Hammer+91}, $^{15}$\citet{Carrasco+17},$^{16}$\citet{Chan+19},$^{17}$\citet{Tyson+90},$^{18}$\citet{Limousin+09}, $^{19}$\citet{Auger+13}.}
\label{tab:Table3}
\input{tabs/Table3_1.tex}
\end{table*}

\begin{table*}
\ContinuedFloat
\caption{\textit{Continued}.}
\input{tabs/Table3_2.tex}
\end{table*}

\begin{table*}
\ContinuedFloat
\caption{\textit{Continued}.}
\input{tabs/Table3_3.tex}
\end{table*}

\begin{table*}
\ContinuedFloat
\caption{\textit{Continued}.}
\input{tabs/Table3_4.tex}
\end{table*}

We provide the full candidate systems of the SuGOHI-c in the online material\footnotemark[3]. The list of SuGOHI-c with grades A and B is presented in \tref{tab:Table3} which provides the system name, the equatorial coordinates, the lens and source redshift, the arc radius, the mean and the $\sigma$ of the rank, a qualitative grade, the parent catalog, and references from previous studies. \fsref{fig:Figure4} show composite colour ($gri$ or $riz$) cutouts of grades A and B for the SuGOHI-c sample. At the top of each cutout is the system name. At the bottom left is a grade (labeled "A" or "B"), as well as a label "$\mathbb{C}$", "$\mathbb{D}$", or "$\mathbb{K}$" if the lens is spectroscopically confirmed, previously discovered, and well-known, respectively. The known candidates have been identified by cross-matching with the published systems in the literature, as reported in \tref{tab:Table3}.

We show the photometric redshift distribution of lens galaxies and arc radii of the systems in \fref{fig:Figure5}, and for comparison, we also show 125 lens systems of the SARCS sample distribution with R$_{\rm arc}\geq2$ arcsec from \cite{More+12}. We find that the mean lens redshift for the SuGOHI-c sample and SARCS sample are $z=0.50\pm0.23$ and $z=0.58\pm0.22$, respectively. We note that the mean redshifts of both samples have good agreement. We find that the peak of the SuGOHI-c lens sample, on the other hand, is at  $z\sim0.3-0.4$ which is consistent with the peak expected from numerical simulation \citep{Bartelmann+98}.

During our inspection, we also found a number of strong lens systems with red-coloured sources (e.g., HSC J0211$-$0343, HSC J1143$+$0102). We mark such systems with a $\dagger$ in \tref{tab:Table3}. Some of them are high redshift galaxies at $z\sim6$ (Oguri et al., in prep and Ono et al., in prep). We further note that HSC J2211$-$0008 has a spectroscopically confirmed lensed source which is a Lyman-break galaxy at $z=2.26$. Details of the follow-up Subaru observations and analysis of this system will be reported in More et al. (in prep).


\section{Spectroscopic Follow-up} \label{sec:follow}
We carried out spectroscopic observations of 10 candidates from SuGOHI-c sample in order to confirm the lensing nature and obtain spectroscopic redshifts essential for detailed mass modelling of strong lenses (Jaelani et al., in prep.). Our sample was part of the larger spectroscopic campaign for SuGOHI lenses (ESO programme 099.A-0220, PI: S. Suyu) with the Very Large Telescope (VLT)'s X-shooter. These candidates were selected from an early sample of grade A-B lenses with $z_{\ell}>0.6$ from a smaller footprint. X-shooter is an Echelle spectrograph \citep{Vernet+11}, with an allowed wavelength range $\lambda\lambda$3,000 - 25,000 \AA~. The spectra are acquired through three arms, the ultraviolet (UVB, $\lambda\lambda$3,000 - 5,500 \AA), the visual (VIS, $\lambda\lambda$5,000 - 10,500 \AA), and the near-infrared (NIR, $\lambda\lambda$10,000 - 25,000 \AA). The lensed sources were observed using slit widths of 1.0, 0.9, and 0.9 arcsec~ in the UVB, VIS, and NIR arms, respectively, with a binning of 2 $\times$ 2 applied to the UVB and VIS data. We set the position angle (PA) of the long slit to be preferentially along the lensed arc (see \fref{fig:Figure6}). In order to optimise sky background subtraction, we dithered the observations in the standard ABBA nodding pattern. 

Each system was observed in slit mode during either one (e.g., HSC J0224$-$0336) or two (e.g., HSC J1202$+$0039) observation blocks (OBs), to reach the optimal signal to noise (S/N) ratio. Each OB corresponds to roughly one hour of telescope time, and consists of $10\times285$s exposures obtained in an ABBA nodding pattern, to optimise background subtraction in the NIR arm. Exposure times in the UVB and VIS arms are slightly shorter due to the longer readout time. Observations were executed with a seeing FWHM $<0.9$ arcsec on target position. Initially, we reduce the spectroscopic data using the ESO \textsc{Reflex} software (version 2.9.0) combined with X-shooter pipeline recipes (v3.1.0) \citep{Freudling+13,Modigliani+10}. The pipeline recipes performs standard bias subtraction and flat-fielding of the raw spectra. Cosmic rays are removed using \texttt{LACosmic} \citep{Dokkum+01}. For each arm, we extract the orders and rectify them in wavelength space using a wavelength solution previously obtained from the calibration frames. The resulting rectified orders are then shifted, co-added, and flux calibrated to obtain the final two-dimensional (2D) spectrum. For further data processing and analysis, we use standard \textsc{iraf} tools. We produce 1D spectra using an extraction aperture in all three arms and for all three images of the source (apertures are shown by red and blue dashed line on \fref{fig:Figure6}).

\begin{figure*}
\includegraphics[width=0.941\textwidth]{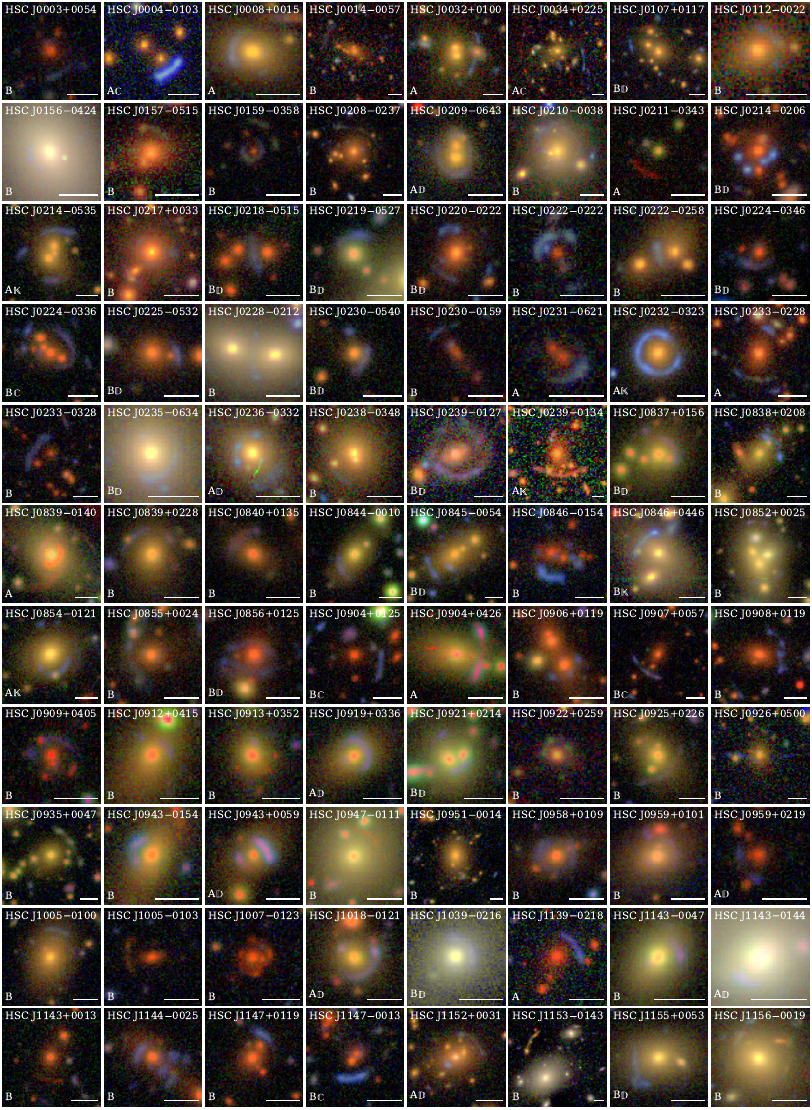}
\caption{SuGOHI-c lens candidates with grades A and B (shown on the bottom left). Spectroscopically confirmed, previously discovered, or known lenses are indicated by "$\mathbb{C}$", "$\mathbb{D}$", or "$\mathbb{K}$", respectively. All images are oriented with North up and East left. Scale bars of 5 arcsec are displayed in the bottom right corner.}
\label{fig:Figure4}
\end{figure*}

\begin{figure*}
\ContinuedFloat
\includegraphics[width=0.941\textwidth]{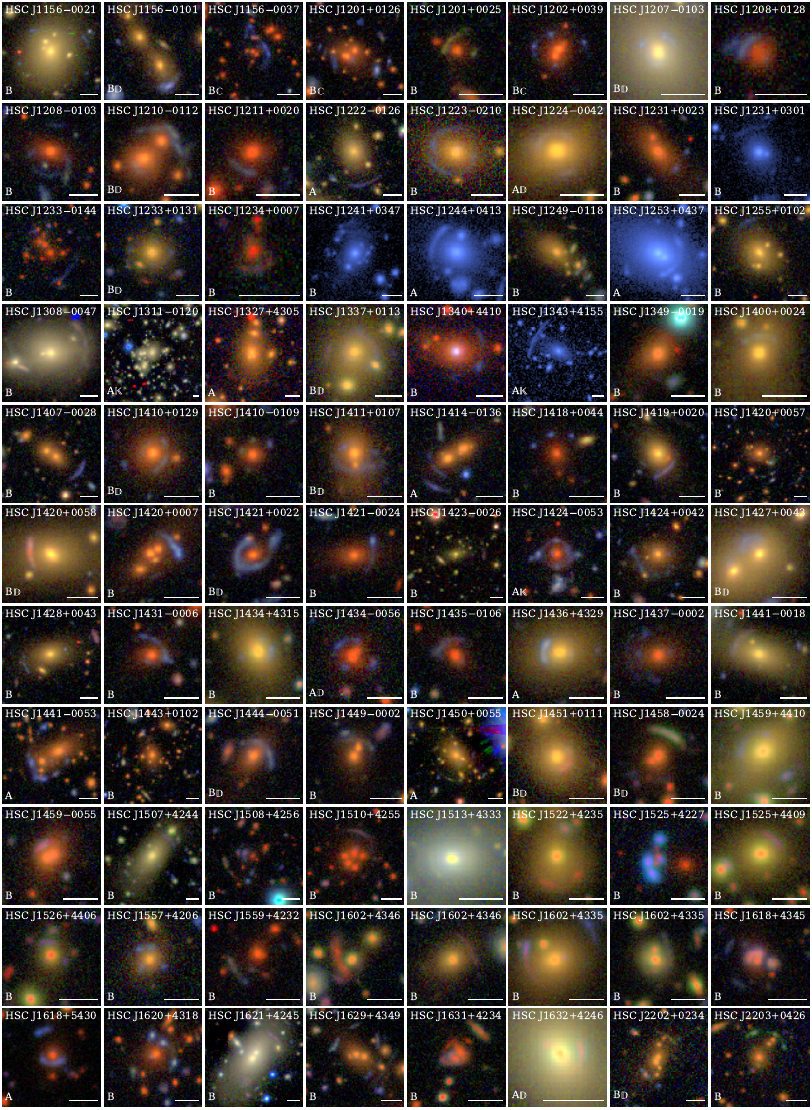}
\caption{\textit{Continued}.}
\end{figure*}

\begin{figure*}
\ContinuedFloat
\includegraphics[width=0.941\textwidth]{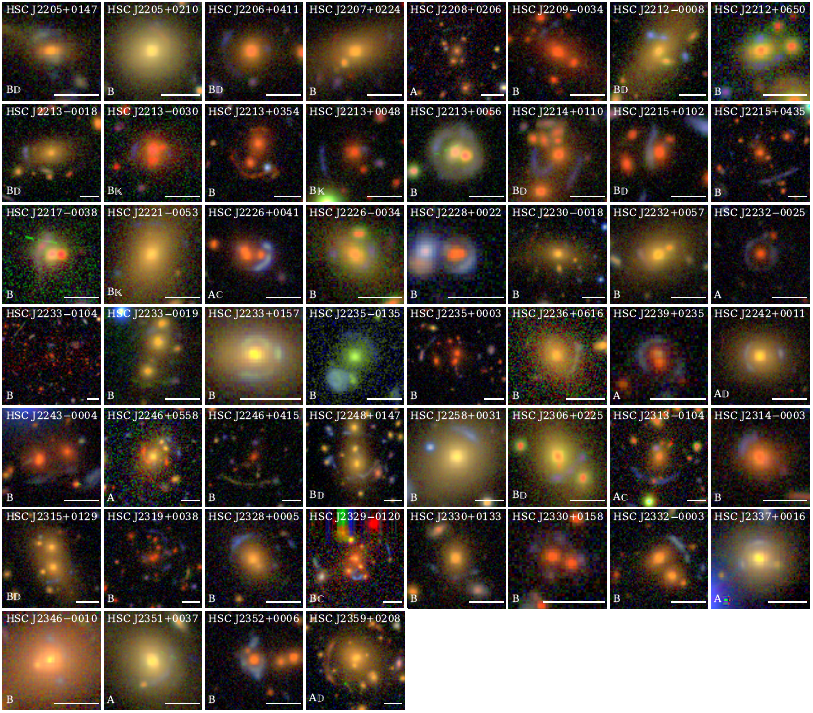}
\caption{\textit{Continued}.}
\end{figure*}

\begin{figure*}
\includegraphics[width=0.90\textwidth]{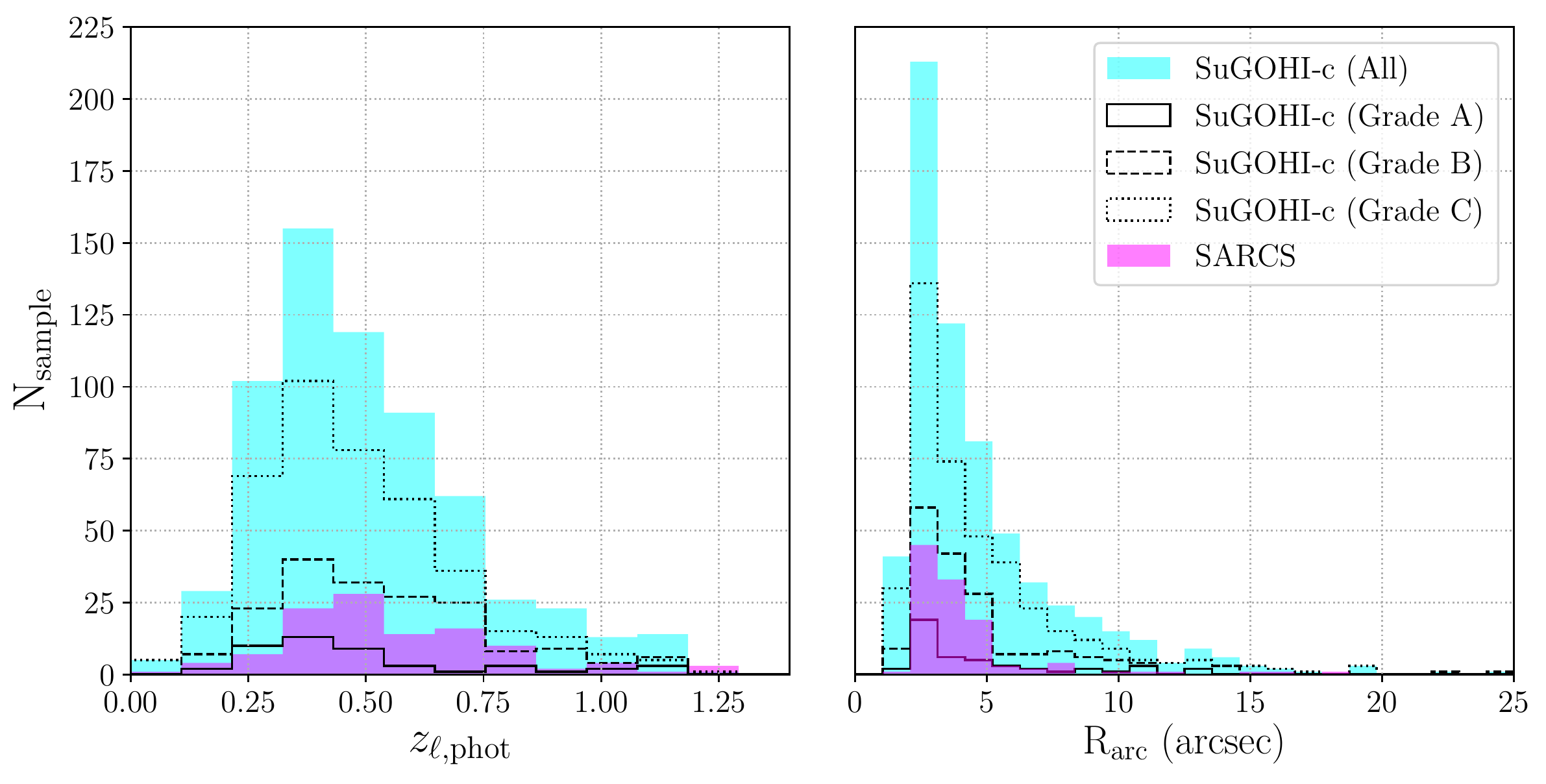}
\caption{\textit{Left:} Photometric redshift distributions of group-to-cluster-scale lens candidates. The peak of the redshift distributions for both the SuGOHI-c sample (cyan) and the SARCS sample \citep[magenta,][]{More+12} are around $z\sim0.4$. \textit{Right:} The binned distribution of arc radii for $R_{\rm arc}\geq2$. The peak at around 3 arcsec attests to the fact that most of the candidates are, indeed, at group-scales. As before, cyan and magenta show the SuGOHI-c and the SARCS samples, respectively.}
\label{fig:Figure5}
\end{figure*}

\begin{figure*}
\includegraphics[width=0.975\textwidth]{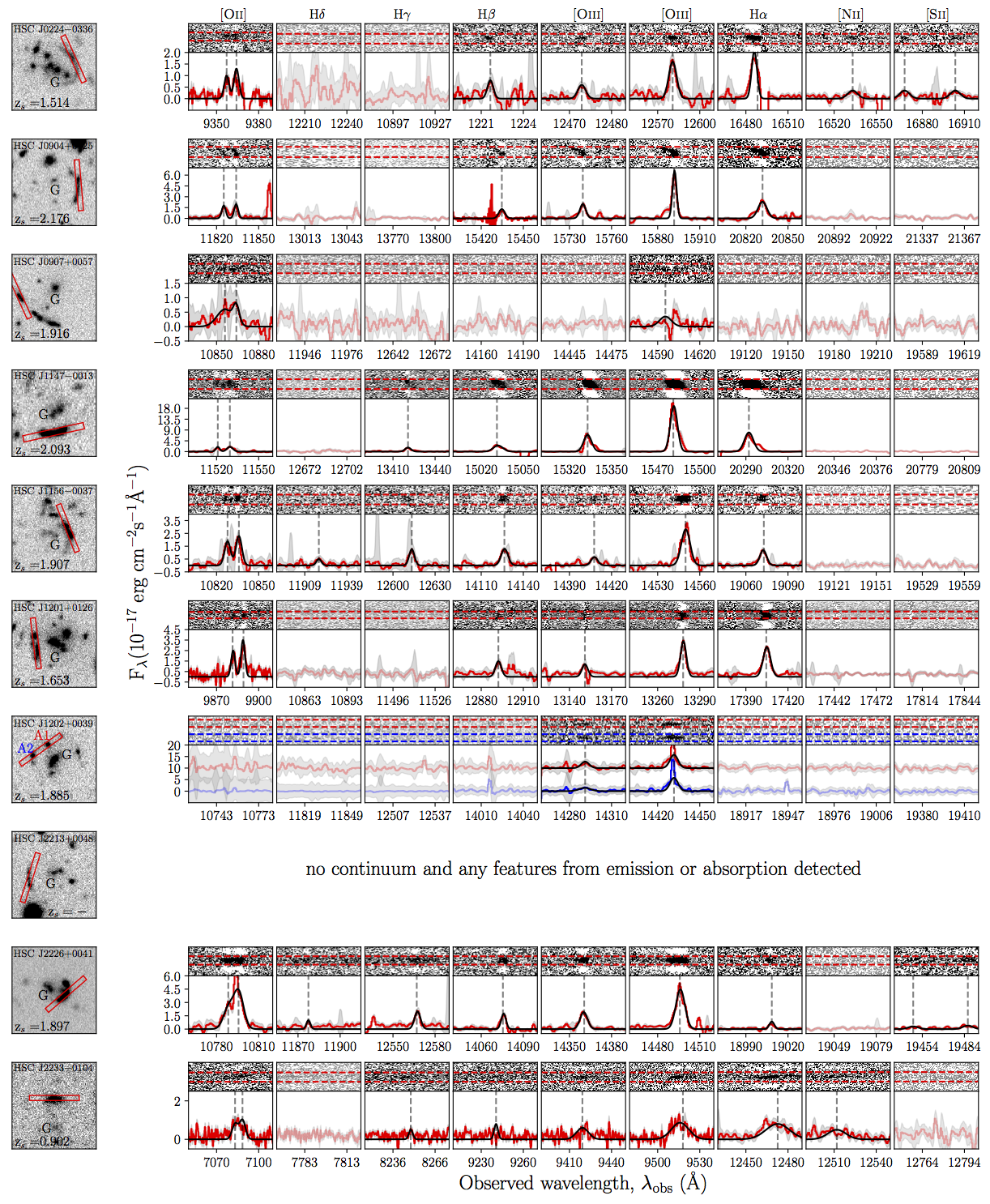}
\caption{Spectroscopic results from the X-shooter observations. Each row is for one candidate lens. One candidate (HSC J2213$+$0048) with no detection is shown image only here. On the left, we show the alignment of the slit on the putative lensed source that was observed. The brightest lens galaxies are denoted by "G". On the right, the upper panel (for each row) shows regions from the 2D spectrum where interesting features are expected. The red (and blue) dashed lines show the exact region used for extracting the 1D spectrum of one of the lensed images (and its counterpart). The bottom panel shows the corresponding stacked 1D spectra (red) where the vertical lines highlight the location of expected emission lines. The labels for those emission lines are shown at the top of the figure. The error on the spectrum is shown by a shaded region (grey) and Gaussian fits to the emission lines are in black. The semi-transparent panels show the locations of some of the common emission lines, which are not detected, for the given source redshift.}
\label{fig:Figure6}
\end{figure*}

\begin{table}
\caption{Summary of X-shooter spectroscopic observations. Position angles (P.A.) are measured East of North.}\label{tab:Table4}
\input{tabs/Table4.tex}
\end{table}

\subsection{Redshift measurement}
We visually inspected all of the three arms of the X-shooter (2D and 1D) spectra in order to identify any emission and/or absorption lines arising from the lensed galaxies. We identified a set of emission lines that could be attributed to a common redshift. We fitted Gaussian profiles to each of those lines in order to determine their central wavelength and, thereby, determined a mean redshift for each lensed galaxy. Most of the lensed arcs showed [\textsc{Oii}] doublet $\lambda3726.03$, $3728.81$ \AA, H$\delta$ $\lambda4101.73$ \AA, H$\gamma$ $\lambda4340.46$ \AA, H$\beta$ $\lambda 4861.32$ \AA, [\textsc{Oiii}] $\lambda4958.91$, $5006.84$ \AA, H$\alpha$ $\lambda6562.79$ \AA, [\textsc{Nii}] $\lambda6583.45$ \AA, [\textsc{Sii}] $\lambda6716.43$, and $\lambda6730.81$ \AA ~which are expected to be found in blue star forming galaxies. Our lensed galaxies span a redshift range from $z\sim0.9$ to $2.2$ (summarised in \tref{tab:Table4}). We give a short description of the confirmed lenses below.

\subsection{Spectroscopically confirmed group-scale lenses}
\textit{HSC J0224$-$0336} at $\zls=0.613$: This system has been reported in \citet{More+12} and has four bright early type galaxies at the centre, surrounded by a blue arc (almost complete ring). We set the slit along the arc to the North-West of the lens (see \fref{fig:Figure6}) that has a peak flux in $g$-band. Most of the emission lines in this system are detected in the NIR arm of X-shooter, H$\beta$, [\textsc{Oiii}], H$\alpha$, [\textsc{Nii}] and [\textsc{Sii}]. We also detected the [\textsc{Oii}] doublet in the VIS arm. These emission lines correspond to a mean redshift of $z=1.514$. The lens galaxy, "G" of HSC J0224$-$0336 shown in \fref{fig:Figure6}, is identified as the centre of the cluster by \citet{Wen+12} which has a richness of $N_{\rm ric, \textsc{Wen}}= 21.62$ with 14 member galaxies and corresponds to $M_{200}\approx 1.18\times 10^{14} \Msun$. However, the same galaxy is identified as a member galaxy of a cluster, in \textsc{camira}, with a richness of $N_{\rm ric, \textsc{camira}}= 19.62$, 63 member galaxies and $M_{200}\approx 6.36\times 10^{13} \Msun$. The stellar velocity dispersion of the lens galaxy is $448 \pm 101$ \kms from the SDSS data. \medskip

\noindent \textit{HSCJ0904$+$0125} at $\zlp=0.914$: For this system, we set the slit along a nearly north-south blue arc. We detected emission lines such as the [\textsc{Oii}] doublet, H$\beta$, [\textsc{Oiii}], and H$\alpha$ in NIR arm corresponding to a mean redshift of $z=2.176$. The lens galaxy of the system, "G" in HSCJ0904$+$0125 panel of \fref{fig:Figure6}, is identified as the galaxy member in \textsc{camira}. The cluster has a richness $N_{\rm ric, \textsc{camira}}= 18.37$ with 56 member galaxies and corresponds to $M_{200}\approx 5.83\times 10^{13} \Msun$.\medskip

\noindent \textit{HSC J0907$+$0057} at $\zlp=0.723$: This system is composed of a number of blue arcs around a bright early type galaxy at a separation $\simeq5$ arcsec. We set a slit along the east-most arc-like component whose light is not contaminated by any red blobs (see \fref{fig:Figure4}). We detect weak emission lines such as the [\textsc{Oii}] doublet and [\textsc{Oiii}]$\ \lambda5008.24$ \AA ~line, yielding a lensed galaxy redshift of $z=1.916$. The lens galaxy, "G" of HSC J0907$+$0057 shown in \fref{fig:Figure6}, is at the centre of the cluster in \textsc{camira} which has a richness of $N_{\rm ric, \textsc{camira}}= 35.49$ and 75 member galaxies corresponding to $M_{200}\approx 1.38\times 10^{14} \Msun$.\medskip

\noindent \textit{HSC J1147$-$0013} at $\zlp=0.805$: We detect many strong emission lines such as [\textsc{Oii}] doublet, H$\gamma$, H$\beta$, [\textsc{Oiii}] and H$\alpha$ in the NIR arm from the nearly straight blue arcs. This system has similar features to HSC J0904+0125 which has a small peak near main peak. We find that the emission lines correspond to a mean redshift of $z=2.093$. This group-scale system is found serendipitously during the inspection owing to the very bright arc next to another cluster. The cluster catalogs may have missed this due to lack of sufficiently bright galaxies. \medskip

\noindent \textit{HSC J1156$-$0037} at $\zlp=0.918$: We find that the blue arc has a mean redshift of $z=1.907$ from the emission lines [\textsc{Oii}] doublet, H$\gamma$, H$\beta$, [\textsc{Oiii}] and H$\alpha$ in the NIR arm. The lens galaxy, "G" of HSC J1156$-$0037 shown in \fref{fig:Figure6}, is found to be a galaxy member of the large cluster in \textsc{camira} which has a high richness of $N_{\rm ric, \textsc{camira}}= 64.05$ corresponding to $M_{200}\approx 3.00\times 10^{14} \Msun$. This cluster has 126 member galaxies.\medskip

\noindent \textit{HSC J1201$+$0126} at $\zlp=0.618$: This system has been reported in \citet{Petrillo+19}. We set the slit along the blue arc which has a small early type galaxy included (which produces the continuum in the 2D spectra). We detect weak continuum from early type galaxy and strong emission lines, H$\beta$, [\textsc{Oiii}] and H$\alpha$ in NIR arm and [\textsc{Oii}] doublet in VIS arm, yielding a lensed galaxy redshift of $z=1.653$. The lens galaxy, "G" of HSC J1201$+$0126 shown in \fref{fig:Figure6}, is at the centre of the cluster as per \textsc{camira} which has a richness of $N_{\rm ric, \textsc{camira}}= 36.23$ with 67 member galaxies and corresponds to $M_{200}\approx 1.42\times 10^{14} \Msun$.\medskip

\noindent \textit{HSC J1202$+$0039} at $\zls=0.689$: As seen in \fref{fig:Figure6}, the slit targeting this system covers the source in two locations, the East and North of the lens. We detect strong emission lines [\textsc{Oiii}]$\ \lambda4960.30$ \AA ~and [\textsc{Oiii}]$\ \lambda5008.24$ \AA ~in each source location. We find that the emission lines have a mean redshift of $z=1.885$. The lens galaxy, "G" of HSC J1202$+$0039 shown in \fref{fig:Figure6}, is identified as the BCG as per \textsc{camira} which has a richness of $N_{\rm ric, \textsc{camira}}= 36.298$ with 75 member galaxies and corresponding to $M_{200}\approx 1.42\times 10^{14} \Msun$. The stellar velocity dispersion of the BCG is $238 \pm 37$ \kms from the SDSS data.\medskip

\noindent \textit{HSC J2213$+$0048} at $\zlp=0.945$: We do not detect continuum or any features from emission or absorption in the spectrum of this system. The lens galaxy, "G" of HSC J2213$+$0048 shown in \fref{fig:Figure6}, is identified as a member of a group in \citet{Ford+15} which has a richness of $N_{\rm ric, \textsc{Ford}}= 8.80$ with 68 member galaxies, corresponding to $M_{200}\approx 7.29\times 10^{12} \Msun$. This system also has been reported in \citet{More+12}.\medskip

\begin{figure}
\centering
\includegraphics[width=0.40\textwidth]{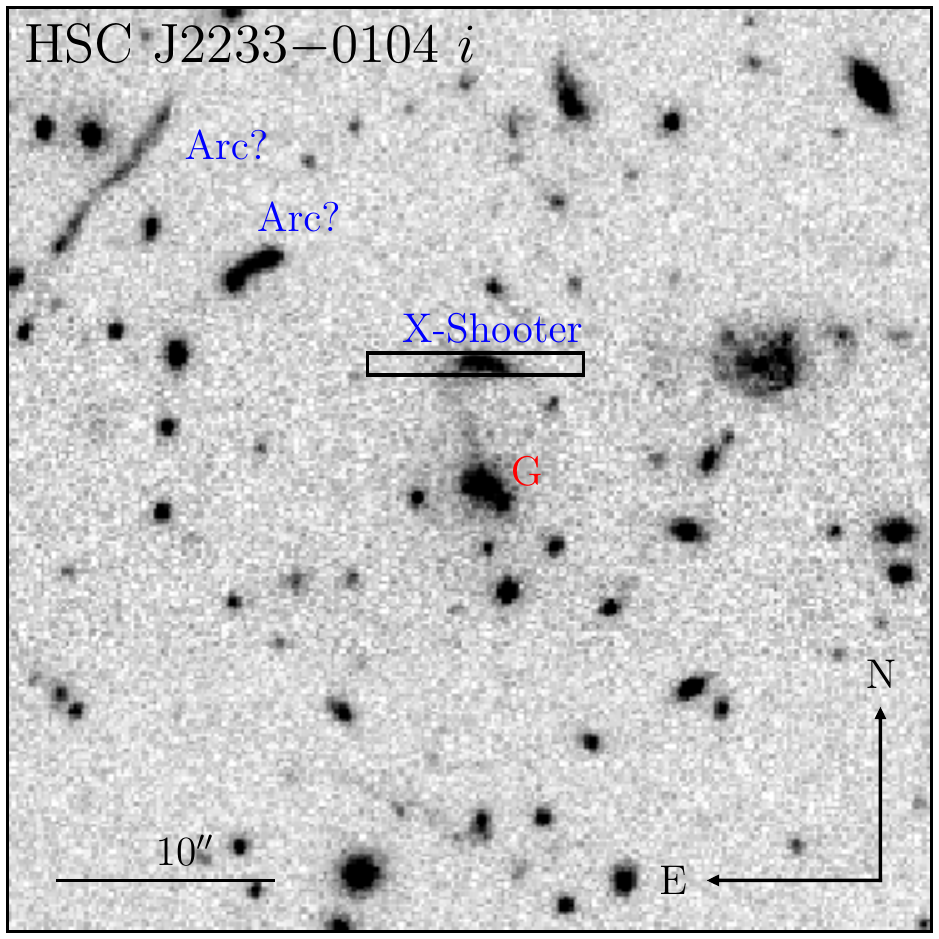}
\caption{HSC J2233$-$0104 lens candidate. Image is $\sim43$ arcsec on the side. The bar shows a scale of 10 arcsec.}
\label{fig:Figure7}
\end{figure}

\noindent \textit{HSC J2226$+$0041} at $\zls=0.647$: This is a known lens system \citep{Diehl+17,Sonnenfeld+18,Jacobs+19}. We measure a mean redshift of $z= 1.897$, from emission lines such as [\textsc{Oii}] (assuming the rest-frame centroid of the unresolved [\textsc{Oii}] doublet $\ \lambda3728.3$ \AA), H$\gamma$, H$\beta$, [\textsc{Oiii}]$\ \lambda4960.30$ \AA, [\textsc{Oiii}]$\ \lambda5008.24$ \AA, and H$\alpha$. The stellar velocity dispersion of the lens galaxy, "G" of HSC J2226$+$0041 shown in \fref{fig:Figure6}, is $318 \pm 47$ \kms from the SDSS data.\medskip

\noindent \textit{HSC J2233$-$0104} at $\zlp=0.953$: We detect three probable blue arcs: a long-thin arc and a short arc to the North-East of the lens, and a third short arc to the North of the lens (see \fref{fig:Figure7}). We set the slit along the northern arc and detect some emission lines: an unresolved [\textsc{Oii}] doublet $\lambda3728.30$, [\textsc{Oiii}]$\ \lambda4960.30$ \AA, [\textsc{Oiii}]$\ \lambda5008.24$ \AA, H$\alpha$, [\textsc{Nii}] and also weak emission of H$\beta$. The emission lines suggest a mean redshift of $z=0.902$, indicating that the arc is probably not a lensed galaxy since the redshift of this arc is close to the photometric redshift of the lens galaxy, "G" of HSC J2233$-$0104 shown in \fref{fig:Figure6}.


\section{Summary and Conclusion} \label{sec:summary}
We have carried out the largest ever systematic search for strong gravitational lens systems at group-to-cluster-scales. Since the S18A release of the HSC-SSP Survey, covering nearly 1,114 deg$^2$, we have visually inspected 39,435 groups and clusters selected from four parent cluster catalogs. While \textsc{camira} catalog was obtained from HSC imaging, other catalogs \citep{Wen+12,Ford+15,Rykoff+16} came from previous surveys with overlapping footprints.

Our search resulted in a total of 641 lens candidates with 228 highly promising (grade A-B) candidates and 413 plausible (grade C) candidates. Additionally, we report 131 galaxy-scale lens candidates found serendipitously during our search. Most of these are new and are missed from the previously reported SuGOHI-g samples (see \aref{app:sugohig}).

The SuGOHI-c will enable detailed studies of mass distributions in individual systems for even low-mass galaxy groups at low to intermediate redshifts and clusters at very high redshifts. Furthermore, the large sample size will surpass any of the previous statistical studies of group-scale lenses. Finally, we have nearly six times more lenses at high redshifts ($\zl>0.8$) compared to the previous high-redshift SARCS sample. Thus, we will be able to study evolution in the mass distributions at these mass scales for the first time. 

The SuGOHI-c sample has many striking systems with blue giant arcs, red lensed galaxies, and in some cases, multiple lensed galaxies from distinct redshifts lensed by the same galaxy groups. We also present the results of our spectroscopic follow-up with X-shooter where, for 9 out of the 10 candidates, we could detect emission lines and successfully measure the redshifts of the lensed galaxies. A detailed mass modelling analysis using spectroscopic results will be presented in the near future.

\section*{Acknowledgements}
The authors would like to thank the referee for improving the presentation of the paper. ATJ and KTI are supported by JSPS KAKENHI Grant Number JP17H02868. MO is supported by JSPS KAKENHI Grant Number JP15H05892 and JP18K03693. IK is supported by JSPS KAKENHI Grant Number JP15H05896. SHS thanks the Max Planck Society for support through the Max Planck Research Group. J. H. H. C. acknowledges support from the Swiss National Science Foundation (SNSF). This work was supported in part by World Premier International Research centre Initiative (WPI Initiative), MEXT, Japan. The Hyper Suprime-Cam (HSC) collaboration includes the astronomical communities of Japan and Taiwan, and Princeton University. The HSC instrumentation and software were developed by the National Astronomical Observatory of Japan (NAOJ), the Kavli Institute for the Physics and Mathematics of the Universe (Kavli IPMU), the University of Tokyo, the High Energy Accelerator Research Organization (KEK), the Academia Sinica Institute for Astronomy and Astrophysics in Taiwan (ASIAA), and Princeton University. Funding was contributed by the FIRST program from Japanese Cabinet Office, the Ministry of Education, Culture, Sports, Science and Technology (MEXT), the Japan Society for the Promotion of Science (JSPS), Japan Science and Technology Agency (JST), the Toray Science Foundation, NAOJ, Kavli IPMU, KEK, ASIAA, and Princeton University. 

This paper makes use of software developed for the Large Synoptic Survey Telescope. We thank the LSST Project for making their code available as free software at \href{http://dm.lsst.org}{http://dm.lsst.org}. The Pan-STARRS1 Surveys (PS1) have been made possible through contributions of the Institute for Astronomy, the University of Hawaii, the Pan-STARRS Project Office, the Max-Planck Society and its participating institutes, the Max Planck Institute for Astronomy, Heidelberg and the Max Planck Institute for Extraterrestrial Physics, Garching, The Johns Hopkins University, Durham University, the University of Edinburgh, Queen's University Belfast, the Harvard-Smithsonian centre for Astrophysics, the Las Cumbres Observatory Global Telescope Network Incorporated, the National Central University of Taiwan, the Space Telescope Science Institute, the National Aeronautics and Space Administration under Grant No. NNX08AR22G issued through the Planetary Science Division of the NASA Science Mission Directorate, the National Science Foundation under Grant No. AST-1238877, the University of Maryland, and Eotvos Lorand University (ELTE) and the Los Alamos National Laboratory.

Based [in part] on data collected at the Subaru Telescope and retrieved from the HSC data archive system, which is operated by Subaru Telescope and Astronomy Data centre at National Astronomical Observatory of Japan.

Funding for the Sloan Digital Sky Survey IV has been provided by the Alfred P. Sloan Foundation, the U.S. Department of Energy Office of Science, and the Participating Institutions. SDSS acknowledges support and resources from the centre for High-Performance Computing at the University of Utah. The SDSS web site is \href{www.sdss.org}{www.sdss.org}.

SDSS is managed by the Astrophysical Research Consortium for the Participating Institutions of the SDSS Collaboration including the Brazilian Participation Group, the Carnegie Institution for Science, Carnegie Mellon University, the Chilean Participation Group, the French Participation Group, Harvard-Smithsonian centre for Astrophysics, Instituto de Astrof\'{i}sica de Canarias, The Johns Hopkins University, Kavli Institute for the Physics and Mathematics of the Universe (IPMU)/University of Tokyo, the Korean Participation Group, Lawrence Berkeley National Laboratory, Leibniz Institut f\"{u}r Astrophysik Potsdam (AIP), Max-Planck-Institut f\"{u}r Astronomie (MPIA Heidelberg), Max-Planck-Institut f\"{u}r Astrophysik (MPA Garching), Max-Planck-Institut f\"{u}r Extraterrestrische Physik (MPE), National Astronomical Observatories of China, New Mexico State University, New York University, University of Notre Dame, Observat\'{o}rio Nacional/MCTI, The Ohio State University, Pennsylvania State University, Shanghai Astronomical Observatory, United Kingdom Participation Group, Universidad Nacional Aut\'{o}noma de M\'{e}xico, University of Arizona, University of Colorado Boulder, University of Oxford, University of Portsmouth, University of Utah, University of Virginia, University of Washington, University of Wisconsin, Vanderbilt University, and Yale University.


\bibliographystyle{mnras}
\bibliography{references_papers}

\appendix
\section{Additional serendipitous lens candidates from the HSC-SSP S18A}\label{app:sugohig}

During our visual inspection of galaxy groups and clusters, some galaxy-scale lenses were discovered serendipitously which happen to be either member galaxies of the group or field galaxies in the vicinity. Since the lensing is due to an individual galaxy rather than a group/cluster (e.g., see the right most panel of \fref{fig:Figure2}), these systems are excluded from our formal SuGOHI-c sample and are reported here instead.

\begin{figure}
\includegraphics[width=0.45\textwidth]{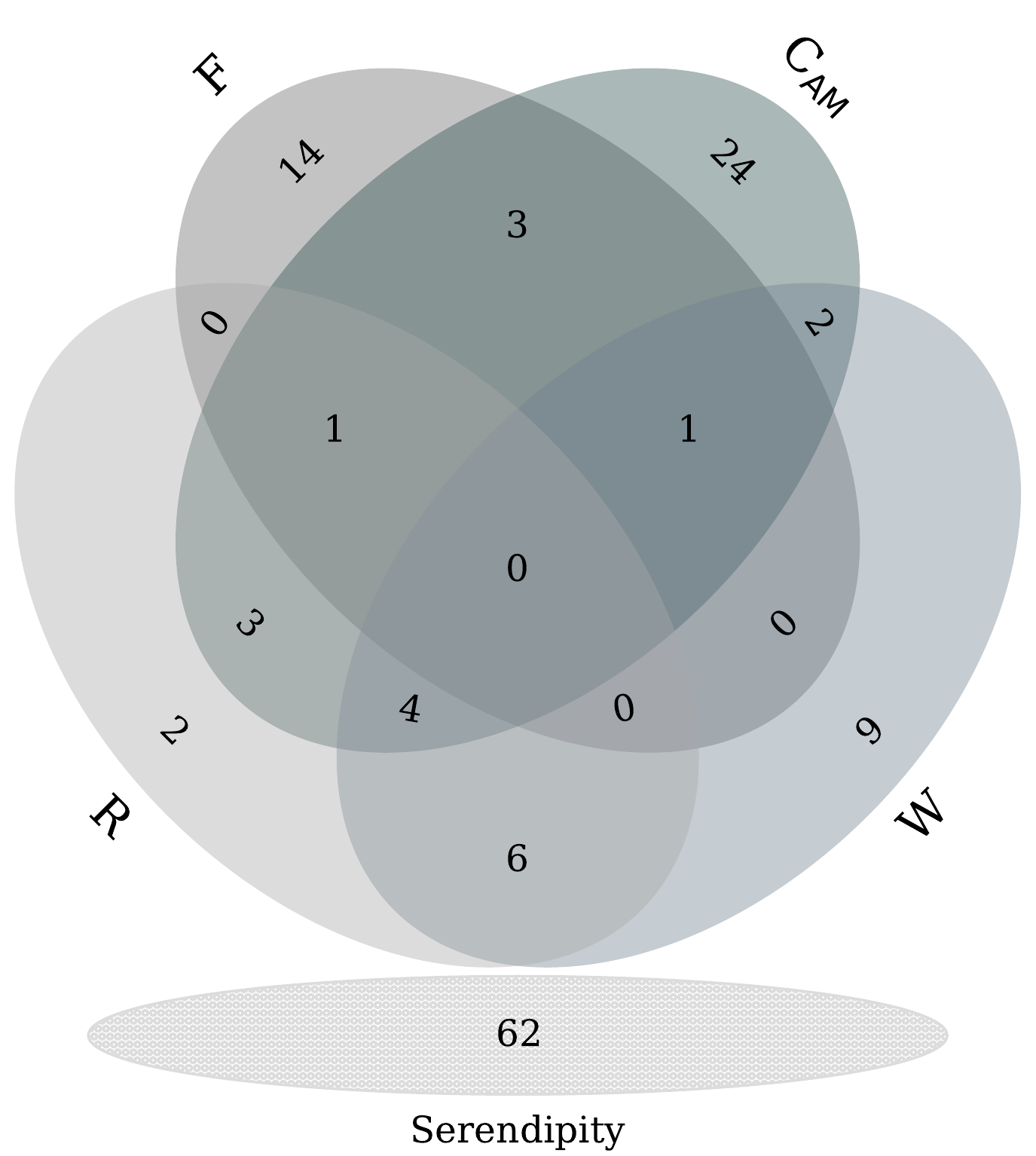}
\caption{Similar to \fref{fig:Figure3}, the distribution of lens candidates at galaxy-scales according to the parent cluster catalogs with R$_{\rm arc}<2$ arcsec. The Venn diagram is divided into two panels: 69 lens systems of SuGOHI-g correspond to the parent cluster (upper) and 62 lens systems which are serendipiously (bottom) discovered during the inspection.}
\label{fig:FigureA1}
\end{figure}

\begin{figure*} 
\centering
\includegraphics[width=0.941\textwidth]{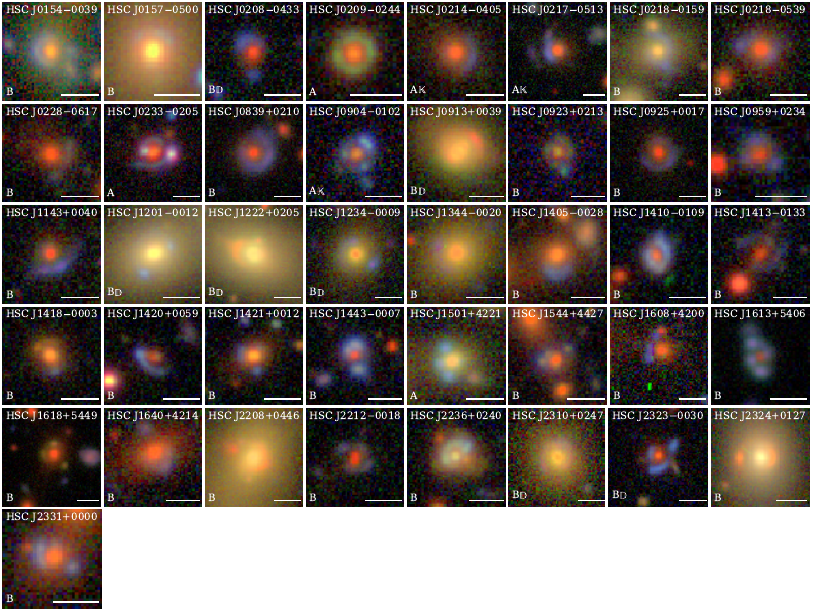}
\caption{Lens candidates at galaxy-scales with grades A and B (labels shown on the bottom left). Spectroscopically confirmed, previously discovered, or known lenses are indicated by "$\mathbb{C}$", "$\mathbb{D}$", or "$\mathbb{K}$", respectively. All images are oriented with North up and East left. A scale bars corresponding to 3 arcsec are displayed in the bottom right corner. The grade C candidates from this sample are made available on the same page as the formal SuGOHI-c sample (\href{http://www-utap.phys.s.u-tokyo.ac.jp/~oguri/sugohi/}{http://www-utap.phys.s.u-tokyo.ac.jp/$\sim$oguri/sugohi/}).}
\label{fig:FigureA2}
\end{figure*}

\begin{table}
\caption{Lens candidate statistics for SuGOHI-g, similar with \tref{tab:Table2}.}
\label{tab:TableA1}
\input{tabs/TableA1.tex}
\end{table}

\begin{table*}
\caption{Extra lens candidates at galaxy-scales discovered serendipitously. References: $^3$\citet{Jacobs+19}, $^4$\citet{Petrillo+19}, $^5$\citet{Sonnenfeld+18}, $^6$\citet{Wong+18}, $^7$\citet{More+12}, $^8$\citet{More+16b}, $^9$\citet{Cabanac+07}, $^{20}$\citet{Jaelani+19}.}
\label{tab:TableA2}
\input{tabs/TableA2.tex}
\end{table*}

\balance
\bsp	
\label{lastpage}
\end{document}

%% file: macros.tex
\newcommand{\sref}[1]{Section~\ref{#1}}
\newcommand{\aref}[1]{Appendix~\ref{#1}}
\newcommand{\fref}[1]{Figure~\ref{#1}}
\newcommand{\fsref}[1]{Figures~\ref{#1}}
\newcommand{\tref}[1]{Table~\ref{#1}}

\newcommand{\kms}{\ifmmode  \,\rm km\,s^{-1} \else $\,\rm km\,s^{-1}$ \fi }
\newcommand{\Msun}{\ifmmode {M_{\odot}} \else ${M_{\odot}}$ \fi}

\def\zcl{z_{\rm cl}}
\def\zl{z_{\ell}}
\def\zls{z_{\ell,\rm spec}}
\def\zlp{z_{\ell,\rm phot}}
\def\zs{z_{\rm s}}

\newcommand\nodata{ ~$\cdots$~ }%

\usepackage[usenames]{color}

%% file: tabs/Table1.tex
\begin{tabular}{L{3.0cm}C{1.9cm}r}
\hline
Catalog          & Catalog label  & Number of Clusters \\
\hline
\textsc{camira}  & \textsc{Cam} & 14,992\\ 
\cite{Ford+15}   & F            & 9,475\\ 
\cite{Rykoff+16} & R            & 2,968\\ 
\cite{Wen+12}    & W            & 12,000 \\ 
\hline
\end{tabular}

%% file: tabs/Table2.tex
\begin{tabular}{L{2cm}ccccC{2.2cm}}
\hline
    & \multicolumn{3}{c}{Grade} & \multirow{2}{*}{Total} & Previously known /\\
    & A & B & C &  & spectroscopically confirmed\\
\hline
SuGOHI-c           &  47 & 181 & 413 & 641 & 105 \\
\quad \textsc{Cam}&  21 & 128 & 202 & 351 &  68 \\
\quad F           &   5 &  15 &  40 &  60 &  20 \\
\quad R           &  20 &  56 & 136 & 212 &  40 \\
\quad W           &  26 &  79 & 194 & 299 &  58 \\
\quad Serendipity &   6 &  24 &  61 &  91 &  12 \\
\hline
\end{tabular}

%% file: tabs/Table3_1.tex
\begin{tabular}{L{2.6cm}rrccccccC{1.8cm}C{1.7cm}}
\hline
Name & $\alpha$(J2000) & $\delta$(J2000) & $\zlp$ & $\zls$ & R$_{\rm arc}$ (arcsec) & Rank & $\rm \sigma_{Rank}$ & Grade & PC & References \\
\hline
HSC J0003$+$0054            &   0.8045 &    0.9069 & 1.13 & \nodata &  3.22 & 1.78 & 0.63 & B & \textsc{Cam}        & \nodata \\
HSC J0004$-$0103\C          &   1.2155 & $-$1.0551 & 0.48 & \nodata &  3.23 & 2.78 & 0.42 & A & \nodata             & $^{1, 15}$\\
HSC J0008$+$0015            &   2.2032 &    0.2641 & 0.40 & 0.397   &  2.57 & 2.75 & 0.46 & A & \nodata             & \nodata \\
HSC J0014$-$0057            &   3.7257 & $-$0.9525 & 0.55 & 0.535   & 13.79 & 2.25 & 0.46 & B & W                   & \nodata \\
HSC J0032$+$0100            &   8.0733 &    1.0102 & 0.37 & 0.390   &  6.48 & 2.88 & 0.35 & A & RW                  & \nodata \\
HSC J0034$+$0225\C          &   8.6173 &    2.4227 & 0.32 & \nodata & 13.01 & 2.75 & 0.46 & A & RW                  & $^{15}$ \\
HSC J0107$+$0117$\dagger$\D &  16.7886 &    1.2918 & 0.44 & 0.422   &  9.64 & 2.00 & 0.47 & B & W                   & $^2$ \\
HSC J0112$-$0022            &  18.1073 & $-$0.3798 & 0.50 & 0.466   &  2.09 & 1.75 & 0.46 & B & \nodata             & \nodata \\
HSC J0156$-$0424            &  29.2265 & $-$4.4071 & 0.12 & \nodata &  3.03 & 2.44 & 0.50 & B & \textsc{Cam}RW      & \nodata \\
HSC J0157$-$0515            &  29.3186 & $-$5.2537 & 0.53 & 0.560   &  2.98 & 2.29 & 0.45 & B & \textsc{Cam}        & \nodata \\   
HSC J0159$-$0358            &  29.9093 & $-$3.9829 & 1.11 & \nodata &  4.38 & 2.29 & 0.45 & B & \textsc{Cam}        & \nodata \\   
HSC J0208$-$0237            &  32.0673 & $-$2.6233 & 0.53 & 0.514   &  7.41 & 2.14 & 0.83 & B & \textsc{Cam}RW      & \nodata \\   
HSC J0209$-$0643\D          &  32.3721 & $-$6.7200 & 0.40 & 0.407   &  3.08 & 2.67 & 0.67 & A & FW                  & $^{3,7}$ \\
HSC J0210$-$0038            &  32.6662 & $-$0.6422 & 0.24 & 0.287   &  5.72 & 1.63 & 0.74 & B & RW                  & \nodata \\
HSC J0211$-$0343$\dagger$   &  32.8150 & $-$3.7299 & 0.75 & \nodata &  3.98 & 3.00 & 0.00 & A & \textsc{Cam}F       & \nodata \\
HSC J0214$-$0206\D          &  33.5333 & $-$2.1081 & 0.67 & \nodata &  3.00 & 2.11 & 0.74 & B & \nodata             & $^3$ \\
HSC J0214$-$0535\K          &  33.5335 & $-$5.5925 & 0.47 & 0.445   &  6.37 & 2.86 & 0.35 & A & \textsc{Cam}FRW     & $^{7,9,18}$ \\ 
HSC J0217$+$0033            &  34.3360 &    0.5536 & 0.36 & 0.381   &  5.04 & 1.88 & 0.84 & B & RW                  & \nodata \\
HSC J0218$-$0515\D          &  34.5306 & $-$5.2601 & 0.56 & 0.649   &  2.61 & 1.57 & 0.90 & B & \textsc{Cam}        & $^{7,9}$ \\
HSC J0219$-$0527\D          &  34.9850 & $-$5.4665 & 0.29 & 0.285   &  3.00 & 2.44 & 0.73 & B & \textsc{Cam}FW      & $^{7,9}$ \\   
HSC J0220$-$0222\D          &  35.1766 & $-$2.3668 & 0.60 & 0.546   &  3.99 & 2.14 & 0.64 & B & \textsc{Cam}        & $^5$ \\  
HSC J0222$-$0222            &  35.5932 & $-$2.3699 & 1.18 & \nodata &  3.18 & 1.67 & 0.67 & B & \textsc{Cam}        & \nodata \\
HSC J0222$-$0258            &  35.7480 & $-$2.9743 & 0.46 & \nodata &  2.12 & 1.57 & 0.50 & B & \textsc{Cam}        & \nodata \\   
HSC J0224$-$0346\D          &  36.0040 & $-$3.7738 & 0.95 & \nodata &  2.56 & 1.86 & 0.64 & B & \textsc{Cam}        & $^7$ \\   
HSC J0224$-$0336$^{\rm X}$\C&  36.0437 & $-$3.6015 & 0.61 & 0.613   &  3.89 & 2.29 & 0.45 & B & \textsc{Cam}W       & $^5$ \\ 
HSC J0225$-$0532\D          &  36.3888 & $-$5.5346 & 0.58 & 0.566   &  3.98 & 1.56 & 0.73 & B & \textsc{Cam}F       & $^8$ \\
HSC J0228$-$0212            &  37.0118 & $-$2.2005 & 0.23 & 0.206   &  3.76 & 2.29 & 0.45 & B & \textsc{Cam}RW      & \nodata \\
HSC J0230$-$0540\D          &  37.5355 & $-$5.6774 & 0.46 & 0.498   &  2.26 & 1.57 & 0.50 & B & \textsc{Cam}F       & $^8$ \\
HSC J0230$-$0159            &  37.7006 & $-$1.9841 & 1.10 & \nodata &  4.50 & 1.56 & 0.68 & B & \textsc{Cam}        & \nodata \\  
HSC J0231$-$0621            &  37.7516 & $-$6.3612 & 1.17 & \nodata &  2.18 & 3.00 & 0.00 & A & \textsc{Cam}        & \nodata \\
HSC J0232$-$0323\K          &  38.2078 & $-$3.3905 & 0.46 & 0.450   &  3.77 & 3.00 & 0.00 & A & \textsc{Cam}        & $^{1,2,3,10}$ \\
HSC J0233$-$0228            &  38.2734 & $-$2.4769 & 0.61 & 0.572   &  4.99 & 2.71 & 0.45 & A & \textsc{Cam}        & \nodata \\    
HSC J0233$-$0328            &  38.3837 & $-$3.4671 & 1.12 & \nodata &  4.17 & 1.71 & 1.03 & B & \textsc{Cam}        & \nodata \\   
HSC J0235$-$0634\D          &  38.9092 & $-$6.5684 & 0.20 & 0.181   &  2.13 & 2.00 & 0.00 & B & \textsc{Cam}W       & $^5$ \\
HSC J0236$-$0332\D          &  39.1554 & $-$3.5389 & 0.28 & 0.269   &  2.12 & 3.00 & 0.00 & A & \textsc{Cam}RW      & $^5$ \\
HSC J0238$-$0348            &  39.5988 & $-$3.8036 & 0.29 & 0.322   &  4.29 & 1.63 & 0.52 & B & RW                  & \nodata \\
HSC J0239$-$0127\D          &  39.9260 & $-$1.4632 & 0.35 & \nodata &  4.26 & 2.33 & 0.47 & B & \nodata             & $^2$ \\
HSC J0239$-$0134\K          &  39.9714 & $-$1.5827 & 0.37 & 0.373   & 10.86 & 2.88 & 0.35 & A & RW                  & $^{3,14}$ \\
HSC J0837$+$0156\D          & 129.3593 &    1.9441 & 0.39 & 0.396   &  2.67 & 2.29 & 0.45 & B & \textsc{Cam}R       & $^{4,5}$ \\   
HSC J0838$+$0208            & 129.7372 &    2.1474 & 0.36 & 0.360   &  6.95 & 1.56 & 0.54 & B & \textsc{Cam}RW      & \nodata \\
HSC J0839$-$0140$\dagger$   & 129.8890 & $-$1.6792 & 0.28 & \nodata &  2.43 & 2.56 & 0.68 & A & W                   & \nodata \\
HSC J0839$+$0228            & 129.9141 &    2.4756 & 0.42 & 0.431   &  3.87 & 1.56 & 0.68 & B & \textsc{Cam}        & \nodata \\  	
HSC J0840$+$0135            & 130.2476 &    1.5970 & 0.56 & 0.550   &  3.42 & 1.56 & 0.50 & B & \textsc{Cam}R       & \nodata \\
HSC J0844$-$0010            & 131.1135 & $-$0.1832 & 0.37 & \nodata &  4.62 & 1.56 & 0.50 & B & \textsc{Cam}RW      & \nodata \\   
HSC J0845$-$0054\D          & 131.3341 & $-$0.9156 & 0.41 & \nodata &  7.70 & 1.71 & 0.70 & B & \textsc{Cam}RW      & $^4$ \\  
HSC J0846$-$0154            & 131.6363 & $-$1.9049 & 1.03 & \nodata &  3.93 & 1.67 & 0.67 & B & \textsc{Cam}        & \nodata \\
HSC J0846$+$0446\K            & 131.6978 &    4.7679 & 0.23 & 0.241   &  4.41 & 1.75 & 0.71 & B & RW                  & $^{10,19}$ \\
HSC J0852$+$0025            & 133.1269 &    0.4203 & 0.29 & \nodata &  9.26 & 2.29 & 0.70 & B & R                   & \nodata \\   
HSC J0854$-$0121\K          & 133.6944 & $-$1.3607 & 0.34 & \nodata &  5.02 & 2.88 & 0.35 & A & FRW                 & $^{4,9,18}$ \\
HSC J0855$+$0024            & 133.9344 &    0.4105 & 0.56 & \nodata &  3.45 & 1.57 & 0.50 & B & \textsc{Cam}RW      & \nodata \\   
HSC J0856$+$0125\D          & 134.0864 &    1.4174 & 0.68 & 0.719   &  2.60 & 2.29 & 0.70 & B & \textsc{Cam}        & $^5$ \\   
HSC J0904$+$0125$^{\rm X}$\C& 136.0180 &    1.4208 & 0.91 & \nodata &  5.20 & 2.29 & 0.45 & B & \textsc{Cam}        & \nodata \\ 
HSC J0904$+$0426		    & 136.1276 &    4.4466 & 0.32 & 0.457   &  4.54 & 3.00 & 0.00 & A & W                   & \nodata \\
HSC J0906$+$0119            & 136.5461 &    1.3308 & 0.65 & 0.605   &  3.82 & 1.67 & 0.47 & B & \nodata             & \nodata \\
HSC J0907$+$0057$^{\rm X}$\C& 136.9767 &    0.9587 & 0.72 & \nodata &  7.00 & 2.29 & 0.70 & B & \textsc{Cam}        & \nodata \\
HSC J0908$+$0119            & 137.0261 &    1.3319 & 0.69 & 0.659   &  5.08 & 1.67 & 0.47 & B & \textsc{Cam}W       & \nodata \\
HSC J0909$+$0405            & 137.2978 &    4.0883 & 0.81 & \nodata &  2.41 & 1.57 & 0.50 & B & \textsc{Cam}        & \nodata \\   
HSC J0912$+$0415            & 138.1252 &    4.2654 & 0.44 & 0.453   &  2.26 & 1.56 & 0.68 & B & \textsc{Cam}W       & \nodata \\
\end{tabular}

%% file: tabs/Table3_2.tex
\begin{tabular}{L{2.6cm}rrccccccC{1.8cm}C{1.7cm}}
\hline
Name & $\alpha$(J2000) & $\delta$(J2000) & $\zlp$ & $\zls$ & R$_{\rm arc}$ (arcsec) & Rank & $\rm \sigma_{Rank}$ & Grade & PC & References \\
\hline
HSC J0913$+$0352$\dagger$   & 138.3040 &    3.8705 & 0.47 & 0.456   &  2.37 & 1.57 & 0.90 & B & \textsc{Cam}R       & \nodata \\
HSC J0919$+$0336\D          & 139.7692 &    3.6107 & 0.45 & 0.444   &  2.15 & 3.00 & 0.00 & A & \textsc{Cam}RW      & $^5$ \\
HSC J0921$+$0214\D          & 140.4025 &    2.2363 & 0.33 & 0.319   &  2.15 & 2.29 & 0.45 & B & \textsc{Cam}W       & $^4$ \\   
HSC J0922$+$0259            & 140.6465 &    2.9950 & 1.10 & \nodata &  2.15 & 2.33 & 0.50 & B & \textsc{Cam}        & \nodata \\   
HSC J0925$+$0226            & 141.2609 &    2.4362 & 0.39 & 0.390   &  4.66 & 1.89 & 0.87 & B & \textsc{Cam}W       & \nodata \\
HSC J0926$+$0500            & 141.6992 &    5.0005 & 0.49 & 0.462   &  2.68 & 1.88 & 0.64 & B & RW                  & \nodata \\
HSC J0935$+$0047            & 143.8137 &    0.7959 & 0.37 & 0.358   &  7.82 & 2.38 & 0.92 & B & \textsc{Cam}R       & \nodata \\
HSC J0943$-$0154            & 145.8653 & $-$1.9149 & 0.44 & 0.450   &  2.42 & 1.56 & 0.50 & B & \textsc{Cam}        & \nodata \\
HSC J0943$+$0059\D          & 145.9510 &    0.9903 & 0.43 & \nodata &  2.43 & 3.00 & 0.00 & A & \textsc{Cam}R       & $^2$ \\
HSC J0947$-$0111$\dagger$   & 146.8682 & $-$1.1925 & 0.26 & 0.239   &  4.08 & 2.00 & 0.47 & B & \textsc{Cam}W       & \nodata \\
HSC J0951$-$0014            & 147.9171 & $-$0.2391 & 0.37 & 0.421   & 14.13 & 1.63 & 0.52 & B & \textsc{Cam}RW      & \nodata \\
HSC J0958$+$0109            & 149.6311 &	1.1603 & 0.57 & 0.550   &  2.72 & 2.11 & 0.60 & B & \textsc{Cam}        & \nodata \\   
HSC J0959$+$0101            & 149.8211 &    1.0329 & 0.45 & 0.446   &  3.08 & 1.67 & 0.47 & B & \textsc{Cam}RW      & \nodata \\
HSC J0959$+$0219\D          & 149.9833 &    2.3169 & 0.97 & \nodata &  2.74 & 2.56 & 0.50 & A & \textsc{Cam}        & $^7$ \\
HSC J1005$-$0100            & 151.3869 & $-$1.0078 & 0.40 & 0.420   &  6.19 & 1.67 & 0.47 & B & \textsc{Cam}W       & \nodata \\
HSC J1005$-$0103            & 151.4756 & $-$1.0660 & 1.05 & \nodata &  3.97 & 1.56 & 0.68 & B & \textsc{Cam}        & \nodata \\
HSC J1007$-$0123$\dagger$   & 151.8009 & $-$1.3879 & 0.93 & \nodata &  2.19 & 1.56 & 0.83 & B & \nodata             & \nodata \\
HSC J1018$-$0121\D          & 154.6972 & $-$1.3591 & 0.39 & 0.388   &  3.21 & 2.88 & 0.35 & A & RW                  & $^2$ \\
HSC J1039$-$0216\D          & 159.7807 & $-$2.2750 & 0.16 & \nodata &  2.05 & 2.11 & 0.74 & B & W                   & $^4$ \\
HSC J1139$-$0218            & 174.8726 & $-$2.3071 & 0.79 & \nodata &  3.52 & 2.78 & 0.42 & A & W                   & \nodata \\
HSC J1143$-$0047            & 175.7553 & $-$0.7867 & 0.28 & \nodata &  2.61 & 1.75 & 0.46 & B & \textsc{Cam}RW      & \nodata \\
HSC J1143$-$0144\D          & 175.8743 & $-$1.7418 & 0.11 & 0.106   &  2.56 & 3.00 & 0.00 & A & RW                  & $^{4,12}$ \\
HSC J1143$+$0013$\dagger$   & 175.9262 &    0.2278 & 0.65 & 0.650   &  5.17 & 2.29 & 0.70 & B & \textsc{Cam}        & \nodata \\   
HSC J1144$-$0025            & 176.1623 & $-$0.4299 & 0.73 & 0.614   &  2.23 & 2.00 & 0.76 & B & \textsc{Cam}W       & \nodata \\   
HSC J1147$+$0119            & 176.7731 &    1.3192 & 0.63 & 0.636   &  3.88 & 1.86 & 0.83 & B & \nodata             & \nodata \\   
HSC J1147$-$0013$^{\rm X}$\C& 176.9383 & $-$0.2307 & 0.81 & \nodata &  3.31 & 2.29 & 0.45 & B & \nodata             & \nodata \\ 
HSC J1152$+$0031\D          & 178.0592 &    0.5239 & 0.46 & 0.466   &  5.43 & 2.71 & 0.45 & A & \textsc{Cam}W       & $^{4,5}$ \\  
HSC J1153$-$0144            & 178.3794 & $-$1.7363 & 0.11 & \nodata & 24.60 & 2.13 & 0.00 & B & \textsc{Cam}RW      & \nodata \\
HSC J1155$+$0053\D          & 178.8366 &    0.8843 & 0.31 & 0.283   &  3.65 & 1.71 & 0.45 & B & \nodata             & $^4$ \\   
HSC J1156$-$0019            & 179.0414 & $-$0.3257 & 0.27 & 0.260   &  4.61 & 1.57 & 0.50 & B & \textsc{Cam}R       & \nodata \\
HSC J1156$-$0021            & 179.0451 & $-$0.3501 & 0.26 & 0.256   &  8.01 & 1.86 & 0.35 & B & \textsc{Cam}RW      & \nodata \\   
HSC J1156$-$0101\D          & 179.0548 & $-$1.0339 & 0.44 & \nodata &  4.60 & 1.57 & 0.50 & B & \textsc{Cam}        & $^4$ \\   
HSC J1156$-$0037$^{\rm X}$\C& 179.2234 & $-$0.6316 & 0.92 & \nodata &  3.86 & 1.71 & 1.03 & B & \textsc{Cam}        & \nodata \\ 
HSC J1201$+$0126$^{\rm X}$\C& 180.2978 &    1.4433 & 0.62 & \nodata &  6.77 & 2.29 & 0.45 & B & \textsc{Cam}        & $^4$ \\ 
HSC J1201$+$0025            & 180.4603 &	0.4222 & 0.88 & \nodata &  1.86 & 1.67 & 0.71 & B & \textsc{Cam}        & \nodata \\   
HSC J1202$+$0039$^{\rm X}$\C& 180.7370 &    0.6584 & 0.70 & 0.689   &  4.18 & 2.29 & 0.45 & B & \textsc{Cam}        & \nodata \\ 
HSC J1207$-$0103\D          & 181.9302 & $-$1.0654 & 0.18 & 0.180   &  2.21 & 2.44 & 0.50 & B & \textsc{Cam}W       & $^4$ \\
HSC J1208$+$0128            & 182.2052 &    1.4791 & 0.71 & \nodata &  2.20 & 2.22 & 0.42 & B & \textsc{Cam}        & \nodata \\
HSC J1208$-$0103            & 182.2306 & $-$1.0512 & 0.71 & 0.662   &  4.73 & 2.43 & 0.50 & B & \textsc{Cam}W       & \nodata \\   
HSC J1210$-$0112\D          & 182.5947 & $-$1.2001 & 0.59 & 0.574   &  4.05 & 2.14 & 0.35 & B & \textsc{Cam}        & $^4$ \\  
HSC J1211$+$0020            & 182.9875 &    0.3482 & 0.79 & \nodata &  2.72 & 1.56 & 0.50 & B & \textsc{Cam}        & \nodata \\
HSC J1222$-$0127            & 185.6074 & $-$1.4485 & 0.29 & 0.295   & 10.76 & 2.78 & 0.42 & A & W                   & \nodata \\
HSC J1223$-$0210            & 185.8964 & $-$2.1754 & 0.35 & 0.439   &  4.39 & 1.89 & 0.57 & B & RW                  & \nodata \\
HSC J1224$-$0042\D          & 186.2094 & $-$0.7044 & 0.40 & 0.403   &  2.41 & 2.56 & 0.50 & A & \textsc{Cam}R       & $^4$ \\
HSC J1231$+$0023            & 187.7810 &    0.3921 & 0.60 & 0.591   &  5.94 & 1.89 & 0.87 & B & \textsc{Cam}W       & \nodata \\
HSC J1231$+$0301            & 187.8949 &    3.0313 & 0.42 & 0.461   &  6.18 & 1.78 & 0.63 & B & W                   & \nodata \\
HSC J1233$-$0144            & 188.2623 & $-$1.7352 & 0.63 & \nodata &  9.72 & 2.00 & 0.67 & B & \textsc{Cam}        & \nodata \\
HSC J1233$+$0131\D          & 188.2803 &    1.5300 & 0.42 & 0.425   &  3.39 & 1.75 & 0.46 & B & \textsc{Cam}RW      & $^4$ \\
HSC J1234$+$0007            & 188.6440 &    0.1190 & 1.04 & \nodata &  2.13 & 1.78 & 0.63 & B & \textsc{Cam}        & \nodata \\
HSC J1241$+$0347            & 190.4916 &    3.7893 & 0.34 & 0.414   &  3.37 & 1.63 & 0.52 & B & R                   & \nodata \\
HSC J1244$+$0413            & 191.2226 &    4.2208 & 0.33 & 0.322   &  4.66 & 2.75 & 0.46 & A & RW                  & \nodata \\
HSC J1249$-$0118            & 192.3172 & $-$1.3085 & 0.48 & \nodata &  9.11 & 1.56 & 0.83 & B & W                   & \nodata \\
HSC J1253$+$0437            & 193.4192 &    4.6181 & 0.22 & 0.243   &  4.60 & 3.00 & 0.00 & A & W                   & \nodata \\
HSC J1255$+$0102            & 193.8525 &    1.0358 & 0.34 & 0.374   & 10.38 & 1.63 & 0.52 & B & RW                  & \nodata \\
HSC J1308$-$0047            & 197.1898 & $-$0.7886 & 0.18 & 0.188   & 11.22 & 2.33 & 0.47 & B & W                   & \nodata \\
HSC J1311$-$0120\K          & 197.8737 & $-$1.3410 & 0.14 & 0.174   & 47.99 & 2.63 & 0.74 & A & RW                  & $^{17}$ \\
HSC J1327$+$4305            & 201.8703 &   43.0833 & 0.36 & 0.374   & 10.78 & 2.75 & 0.46 & A & RW                  & \nodata \\
HSC J1337$+$0112\D          & 204.2834 &    1.2176 & 0.32 & 0.327   &  4.39 & 2.25 & 0.71 & B & RW                  & $^4$ \\
HSC J1340$+$4410            & 205.1241 &   44.1676 & 0.44 & 0.546   &  5.96 & 2.33 & 0.47 & B & \nodata             & \nodata \\
HSC J1343$+$4155\K          & 205.8869 &   41.9175 & 0.40 & 0.418   & 12.83 & 3.00 & 0.00 & A & RW                  & $^{10}$ \\
HSC J1349$-$0019$\dagger$   & 207.2637 & $-$0.3298 & 0.55 & \nodata &  2.91 & 1.56 & 0.68 & B & W                   & \nodata \\
HSC J1400$+$0024            & 210.0880 &    0.4055 & 0.37 & \nodata &  2.28 & 1.86 & 0.64 & B & \textsc{Cam}RW      & \nodata \\
HSC J1407$-$0028            & 211.9735 & $-$0.4715 & 0.49 & 0.471   & 10.32 & 2.29 & 0.70 & B & \textsc{Cam}R       & \nodata \\ 		
HSC J1410$+$0129\D          & 212.5043 &    1.4991 & 0.56 & 0.541   &  2.23 & 2.00 & 0.00 & B & \nodata             & $^{4,6}$ \\
\end{tabular}

%% file: tabs/Table3_3.tex
\begin{tabular}{L{2.6cm}rrccccccC{1.8cm}C{1.7cm}}
\hline
Name & $\alpha$(J2000) & $\delta$(J2000) & $\zlp$ & $\zls$ & R$_{\rm arc}$ (arcsec) & Rank & $\rm \sigma_{Rank}$ & Grade & PC & References \\
\hline
HSC J1410$-$0109            & 212.7087 & $-$1.1607 & 0.65 & \nodata &  2.03 & 1.67 & 0.50 & B & \textsc{Cam}        & \nodata \\
HSC J1411$+$0107\D          & 212.9073 &    1.1223 & 0.48 & 0.462   &  2.57 & 2.29 & 0.70 & B & \textsc{Cam}W       & $^4$ \\
HSC J1414$-$0136            & 213.5874 & $-$1.6128 & 0.53 & 0.511   &  5.52 & 3.00 & 0.00 & A & \nodata             & \nodata \\
HSC J1418$+$0044            & 214.5233 &    0.7431 & 0.92 & \nodata &  3.53 & 1.71 & 1.03 & B & \textsc{Cam}        & \nodata \\   
HSC J1419$+$0020$\dagger$   & 214.7852 &    0.3469 & 0.36 & 0.339   &  4.25 & 2.29 & 0.70 & B & \nodata             & \nodata \\   
HSC J1420$+$0057            & 215.0694 &    0.9547 & 0.49 & 0.507   & 10.96 & 2.29 & 0.70 & B & RW                  & \nodata \\   
HSC J1420$+$0058\D          & 215.0743 &    0.9755 & 0.34 & 0.330   &  3.24 & 2.29 & 0.45 & B & \textsc{Cam}W       & $^4$ \\   
HSC J1420$+$0007            & 215.2019 &    0.1258 & 0.58 & 0.545   &  3.67 & 2.29 & 0.70 & B & \textsc{Cam}        & \nodata \\  
HSC J1421$+$0022\D          & 215.2655 &    0.3720 & 0.67 & \nodata &  2.29 & 2.29 & 0.45 & B & \textsc{Cam}        & $^2$ \\   
HSC J1421$-$0024            & 215.4578 & $-$0.4009 & 0.62 & \nodata &  3.15 & 1.56 & 0.50 & B & \textsc{Cam}        & \nodata \\
HSC J1423$-$0026$\dagger$   & 215.9747 & $-$0.4345 & 0.59 & 0.636   & 14.29 & 2.43 & 0.50 & B & \textsc{Cam}        & \nodata \\   
HSC J1424$-$0053\K          & 216.2042 & $-$0.8892 & 0.87 & 0.795   &  3.33 & 2.86 & 0.35 & A & \textsc{Cam}        & $^{4,5,11}$ \\
HSC J1424$+$0042            & 216.2077 &    0.7004 & 0.47 & 0.477   &  5.29 & 1.57 & 0.50 & B & \textsc{Cam}W       & \nodata \\   
HSC J1427$+$0043\D          & 216.7776 &    0.7207 & 0.28 & 0.295   &  2.63 & 2.00 & 0.76 & B & \textsc{Cam}W       & $^4$ \\
HSC J1428$+$0043            & 217.0751 &    0.7252 & 0.32 & 0.335   &  8.48 & 1.57 & 0.50 & B & \textsc{Cam}RW      & \nodata \\   
HSC J1431$-$0006            & 217.8081 & $-$0.1037 & 0.70 & \nodata &  2.45 & 2.14 & 0.35 & B & \textsc{Cam}        & \nodata \\   
HSC J1434$+$4315            & 218.6586 &   43.2615 & 0.38 & 0.385   &  2.11 & 2.25 & 0.89 & B & RW                  & \nodata \\
HSC J1434$-$0056\D          & 218.7267 & $-$0.9496 & 0.76 & 0.728   &  2.74 & 2.57 & 0.73 & A & \textsc{Cam}        & $^5$ \\  
HSC J1435$-$0106            & 218.8296 & $-$1.1101 & 0.78 & \nodata &  2.69 & 1.78 & 0.67 & B & \textsc{Cam}        & \nodata \\   
HSC J1436$+$4329            & 219.0464 &   43.4891 & 0.39 & 0.386   &  2.70 & 3.00 & 0.00 & A & W                   & \nodata \\
HSC J1437$-$0002            & 219.3929 & $-$0.0488 & 0.63 & 0.627   &  3.62 & 2.29 & 0.45 & B & \textsc{Cam}        & \nodata \\   
HSC J1441$-$0018            & 220.3623 & $-$0.3164 & 0.35 & 0.287   &  5.32 & 2.29 & 0.45 & B & \nodata             & \nodata \\
HSC J1441$-$0053            & 220.3862 & $-$0.8996 & 0.57 & \nodata &  8.93 & 3.00 & 0.00 & A & \textsc{Cam}W       & \nodata \\   
HSC J1443$+$0102$\dagger$   & 220.7898 &    1.0362 & 0.59 & 0.529   & 13.19 & 1.86 & 0.83 & B & \textsc{Cam}W       & \nodata \\   
HSC J1444$-$0051\D          & 221.1198 & $-$0.8617 & 0.53 & 0.575   &  2.04 & 2.29 & 0.88 & B & \nodata             & $^{4,5}$ \\
HSC J1449$-$0002            & 222.4072 & $-$0.0452 & 0.57 & 0.529   &  2.97 & 1.71 & 0.70 & B & \textsc{Cam}        & \nodata \\
HSC J1450$+$0055            & 222.6666 &    0.9279 & 0.44 & 0.421   &  9.95 & 2.86 & 0.35 & A & \nodata             & \nodata \\    
HSC J1451$+$0111\D          & 222.7769 &    1.1927 & 0.37 & 0.391   &  2.12 & 1.57 & 0.50 & B & \nodata             & $^4$ \\
HSC J1458$-$0024\D          & 224.6513 & $-$0.4000 & 0.65 & 0.595   &  3.93 & 2.29 & 0.45 & B & \textsc{Cam}W       & $^4$ \\
HSC J1459$+$4410            & 224.8790 &   44.1802 & 0.32 & 0.323   &  3.03 & 1.57 & 0.35 & B & \textsc{Cam}RW      & \nodata \\
HSC J1459$-$0055            & 224.9882 & $-$0.9230 & 0.60 & 0.939   &  2.22 & 1.56 & 0.68 & B & \textsc{Cam}W       & \nodata \\    
HSC J1507$+$4244            & 226.8466 &   42.7340 & 0.23 & 0.218   & 11.15 & 1.75 & 0.71 & B & \textsc{Cam}RW      & \nodata \\
HSC J1508$+$4256            & 227.1544 &   42.9415 & 0.80 & \nodata &  8.82 & 1.89 & 0.31 & B & \textsc{Cam}        & \nodata \\
HSC J1510$+$4255            & 227.6903 &   42.9324 & 0.75 & \nodata &  7.38 & 2.22 & 0.63 & B & \textsc{Cam}        & \nodata \\
HSC J1513$+$4333            & 228.4882 &   43.5582 & 0.24 & 0.237   &  2.45 & 1.63 & 0.52 & B & \textsc{Cam}RW      & \nodata \\
HSC J1522$+$4235            & 230.6975 &   42.5944 & 0.39 & 0.379   &  4.74 & 1.63 & 0.52 & B & \textsc{Cam}RW      & \nodata \\
HSC J1525$+$4227            & 231.2877 &   42.4642 & 0.85 & \nodata &  2.28 & 2.22 & 0.63 & B & \nodata             & \nodata \\
HSC J1525$+$4409            & 231.4855 &   44.1613 & 0.39 & 0.388   &  3.05 & 1.89 & 0.31 & B & \textsc{Cam}R       & \nodata \\
HSC J1526$+$4406            & 231.6363 &   44.1044 & 0.48 & 0.487   &  2.15 & 1.67 & 0.47 & B & \nodata             & \nodata \\
HSC J1557$+$4206            & 239.3841 &   42.1066 & 0.46 & \nodata &  2.15 & 2.33 & 0.67 & B & \nodata             & \nodata \\
HSC J1559$+$4232            & 239.8367 &   42.5423 & 0.85 & \nodata &  4.37 & 1.56 & 0.50 & B & \textsc{Cam}        & \nodata \\
HSC J1602$+$4346$\dagger$   & 240.5990 &   43.7726 & 0.42 & \nodata &  2.30 & 1.86 & 0.64 & B & \textsc{Cam}        & \nodata \\    
HSC J1602$+$4346            & 240.6045 &   43.7709 & 0.42 & \nodata &  3.38 & 2.29 & 0.45 & B & \textsc{Cam}        & \nodata \\
HSC J1602$+$4335            & 240.7110 &   43.5849 & 0.41 & 0.414   &  4.56 & 2.43 & 0.50 & B & \textsc{Cam}RW      & \nodata \\ 
HSC J1602$+$4335            & 240.7214 &   43.5837 & 0.45 & \nodata &  2.81 & 2.00 & 0.00 & B & \textsc{Cam}RW      & \nodata \\
HSC J1618$+$4345            & 244.5774 &	 43.7574 & 0.72 & 0.899 &  4.69 & 2.11 & 0.78 & B & \nodata             & \nodata \\
HSC J1618$+$5430            & 244.5857 &   54.5052 & 0.79 & \nodata &  2.10 & 2.56 & 0.73 & A & \textsc{Cam}        & \nodata \\    
HSC J1620$+$4318            & 245.1101 &   43.3104 & 0.71 & \nodata &  3.99 & 2.29 & 0.70 & B & \nodata             & \nodata \\    
HSC J1621$+$4245            & 245.3623 &   42.7616 & 0.13 & 0.138   & 13.00 & 1.71 & 0.70 & B & \textsc{Cam}RW      & \nodata \\
HSC J1629$+$4349            & 247.4261 &   43.8280 & 0.55 & 0.528   &  6.33 & 1.71 & 0.70 & B & \textsc{Cam}RW      & \nodata \\
HSC J1631$+$4234$\dagger$   & 247.7866 &   42.5781 & 0.68 & \nodata &  2.12 & 1.57 & 0.90 & B & \textsc{Cam}        & \nodata \\    
HSC J1632$+$4246$\dagger$\D & 248.2406 &   42.7699 & 0.22 & 0.228   &  2.20 & 3.00 & 0.00 & A & W                   & $^{16}$ \\
HSC J2202$+$0234\D          & 330.7369 &    2.5761 & 0.49 & 0.482   &  6.60 & 2.29 & 0.70 & B & \textsc{Cam}FRW     & $^8$ \\    
HSC J2203$+$0426            & 330.9438 &    4.4459 & 0.51 & 0.527   &  8.28 & 2.00 & 0.67 & B & F                   & \nodata \\
HSC J2205$+$0147\D          & 331.2789 &    1.7844 & 0.48 & 0.476   &  2.36 & 1.67 & 0.82 & B & F                   & $^7$ \\
HSC J2205$+$0210$\dagger$   & 331.3976 &    2.1760 & 0.26 & 0.252   &  3.29 & 1.56 & 0.68 & B & W                   & \nodata \\
HSC J2206$+$0411\D          & 331.6751 &    4.1919 & 0.53 & 0.537   &  4.24 & 1.56 & 0.68 & B & FW                  & $^7$ \\
HSC J2207$+$0224$\dagger$   & 331.8298 &    2.4046 & 0.42 & 0.418   &  3.50 & 2.29 & 0.70 & B & \textsc{Cam}F       & \nodata \\
HSC J2208$+$0206            & 332.2499 &    2.1152 & 1.04 & \nodata &  5.83 & 2.57 & 0.73 & A & \textsc{Cam}F       & \nodata \\    
HSC J2209$-$0034            & 332.4829 & $-$0.5764 & 0.69 & 0.716   &  4.06 & 2.00 & 0.93 & B & \textsc{Cam}F       & \nodata \\
HSC J2212$-$0008\D          & 333.0476 & $-$0.1389 & 0.36 & 0.365   &  3.40 & 2.29 & 0.70 & B & \textsc{Cam}FRW     & $^1$ \\ 
HSC J2212$+$0650            & 333.1505 &    6.8415 & 0.38 & 0.399   &  2.04 & 2.00 & 0.00 & B & RW                  & \nodata \\
HSC J2213$-$0018\D          & 333.2770 & $-$0.3084 & 0.40 & 0.408   &  7.60 & 1.57 & 0.50 & B & \textsc{Cam}FRW     & $^1$ \\    
HSC J2213$-$0030\K          & 333.2789 & $-$0.5103 & 0.64 & 0.702   &  2.75 & 1.57 & 0.90 & B & \textsc{Cam}F       & $^{7,18}$ \\    
\end{tabular}

%% file: tabs/Table3_4.tex
\begin{tabular}{L{2.6cm}rrccccccC{1.8cm}C{1.7cm}}
\hline
Name & $\alpha$(J2000) & $\delta$(J2000) & $\zlp$ & $\zls$ & R$_{\rm arc}$ (arcsec) & Rank & $\rm \sigma_{Rank}$ & Grade & PC & References \\
\hline
HSC J2213$+$0354$\dagger$   & 333.3342 &    3.9100 & 0.69 & 0.670   &  5.01 & 2.44 & 0.50 & B & \textsc{Cam}        & \nodata \\
HSC J2213$+$0048$^{\rm X}$\K& 333.3826 &    0.8100 & 0.95 & \nodata &  5.19 & 2.29 & 0.45 & B & F                   & $^{7,18}$ \\    
HSC J2213$+$0056            & 333.4550 &    0.9475 & 0.28 & \nodata &  3.14 & 1.57 & 0.50 & B & \textsc{Cam}        & \nodata \\    
HSC J2214$+$0110\D          & 333.5787 &    1.1772 & 0.63 & 0.566   &  3.57 & 1.57 & 0.90 & B & \textsc{Cam}FW      & $^{1,7}$ \\   
HSC J2215$+$0102\D          & 333.8056 &    1.0446 & 0.71 & \nodata &  2.26 & 1.89 & 0.60 & B & \textsc{Cam}F       & $^1$ \\    
HSC J2215$+$0435            & 333.9658 &    4.5838 & 0.65 & \nodata &  9.39 & 2.22 & 0.63 & B & \textsc{Cam}W       & \nodata \\
HSC J2217$-$0038            & 334.3723 & $-$0.6436 & 0.30 & \nodata &  2.06 & 1.56 & 0.50 & B & \nodata             & \nodata \\    
HSC J2221$-$0053\K          & 335.4324 & $-$0.8842 & 0.34 & 0.334   &  4.98 & 1.78 & 0.42 & B & \textsc{Cam}        & $^{18}$ \\
HSC J2226$+$0041$^{\rm X}$\C& 336.5386 &    0.6949 & 0.63 & 0.647   &  2.98 & 3.00 & 0.00 & A & \nodata             & $^{1,3,5}$ \\ 
HSC J2226$-$0034            & 336.6597 & $-$0.5805 & 0.38 & 0.404   &  2.20 & 1.78 & 0.42 & B & \textsc{Cam}R       & \nodata \\
HSC J2228$+$0022            & 337.1687 &    0.3704 & 0.59 & \nodata &  1.21 & 2.00 & 0.54 & B & \nodata             & \nodata \\    
HSC J2230$-$0018$\dagger$   & 337.5731 & $-$0.3125 & 0.40 & 0.406   &  7.05 & 1.57 & 1.05 & B & \textsc{Cam}R       & \nodata \\    
HSC J2232$+$0057            & 338.0466 &    0.9501 & 0.40 & 0.401   &  2.38 & 1.71 & 0.45 & B & \textsc{Cam}W       & \nodata \\
HSC J2232$-$0025            & 338.1611 & $-$0.4261 & 1.08 & \nodata &  2.13 & 3.00 & 0.00 & A & \textsc{Cam}        & \nodata \\
HSC J2233$-$0104$^{\rm X}$  & 338.3201 & $-$1.0694 & 0.95 & \nodata & 22.15 & 1.57 & 1.05 & B & \nodata             & \nodata \\ 
HSC J2233$-$0019            & 338.3331 & $-$0.3264 & 0.45 & 0.398   &  4.13 & 1.57 & 0.73 & B & \textsc{Cam}R       & \nodata \\    
HSC J2233$+$0157            & 338.4742 &    1.9560 & 0.27 & \nodata &  2.08 & 2.22 & 0.63 & B & \textsc{Cam}R       & \nodata \\
HSC J2235$-$0135            & 338.8841 & $-$1.5944 & 0.48 & \nodata &  3.05 & 1.56 & 0.96 & B & W                   & \nodata \\
HSC J2235$+$0003            & 338.9535 &    0.0509 & 0.76 & 0.735   &  8.66 & 1.71 & 0.70 & B & \textsc{Cam}        & \nodata \\   
HSC J2236$+$0616            & 339.0586 &    6.2723 & 0.37 & 0.350   &  3.40 & 2.13 & 0.64 & B & W                   & \nodata \\
HSC J2239$+$0235            & 339.8946 &    2.5853 & 1.13 & \nodata &  1.91 & 3.00 & 0.00 & A & \textsc{Cam}        & \nodata \\
HSC J2242$+$0011\D          & 340.5899 &    0.1956 & 0.39 & 0.385   &  2.43 & 3.00 & 0.00 & A & \textsc{Cam}R       & $^5$ \\   
HSC J2243$-$0004            & 340.9990 & $-$0.0803 & 0.71 & 0.690   &  3.31 & 1.71 & 0.00 & B & \nodata             & \nodata \\   
HSC J2246$+$0558$\dagger$   & 341.5610 &    5.9748 & 0.31 & 0.340   &  2.66 & 2.63 & 0.74 & A & RW                  & \nodata \\
HSC J2246$+$0415		    & 341.6871 &    4.2637 & 1.02 & \nodata &  8.67 & 2.33 & 0.47 & B & \textsc{Cam}        & \nodata \\
HSC J2248$+$0147\D          & 342.2457 &    1.7865 & 0.38 & 0.360   &  6.74 & 2.00 & 0.00 & B & \textsc{Cam}RW      & $^5$ \\
HSC J2258$+$0031            & 344.5655 &    0.5248 & 0.26 & 0.256   &  4.80 & 1.75 & 0.46 & B & \textsc{Cam}RW      & \nodata \\
HSC J2306$+$0225\D          & 346.7428 &    2.4286 & 0.35 & 0.362   &  3.17 & 2.00 & 0.00 & B & \textsc{Cam}RW      & $^6$ \\
HSC J2313$-$0104\C          & 348.4770 & $-$1.0802 & 0.53 & 0.531   &  8.13 & 2.63 & 0.52 & A & \nodata             & $^{1, 15}$ \\
HSC J2314$-$0003            & 348.5673 & $-$0.0529 & 0.60 & \nodata &  2.64 & 1.56 & 0.68 & B & W                   & \nodata \\
HSC J2315$+$0129\D          & 348.9799 &    1.4850 & 0.46 & 0.424   &  3.68 & 1.56 & 0.83 & B & \textsc{Cam}RW      & $^6$ \\
HSC J2319$+$0038            & 349.9726 &    0.6369 & 0.94 & \nodata &  7.69 & 2.33 & 0.82 & B & \textsc{Cam}        & \nodata \\
HSC J2328$+$0005            & 352.2238 &    0.0937 & 0.50 & 0.443   &  3.67 & 2.00 & 0.00 & B & \textsc{Cam}        & \nodata \\  
HSC J2329$-$0120\C          & 352.4494 & $-$1.3466 & 0.53 & 0.537   & 10.50 & 1.67 & 0.94 & B & \textsc{Cam}W       & $^{1, 15}$ \\
HSC J2330$+$0133            & 352.5252 &    1.5512 & 0.42 & 0.444   &  3.44 & 1.67 & 0.67 & B & \textsc{Cam}W       & \nodata \\
HSC J2330$+$0158            & 352.6815 &    1.9702 & 0.69 & \nodata &  2.16 & 1.67 & 0.82 & B & \textsc{Cam}        & \nodata \\
HSC J2332$-$0003            & 353.1491 & $-$0.0511 & 0.52 & 0.510   &  3.76 & 1.71 & 0.45 & B & \textsc{Cam}R       & \nodata \\
HSC J2337$+$0016            & 354.4175 &    0.2781 & 0.32 & 0.272   &  2.00 & 3.00 & 0.00 & A & R                   & \nodata \\
HSC J2346$-$0010            & 356.5148 & $-$0.1829 & 0.26 & 0.261   &  2.30 & 1.56 & 0.68 & B & RW                  & \nodata \\
HSC J2351$+$0037            & 357.8388 &    0.6169 & 0.26 & 0.277   &  2.96 & 2.63 & 0.52 & A & \textsc{Cam}RW      & \nodata \\
HSC J2352$+$0006            & 358.0488 &    0.1041 & 0.67 & \nodata &  1.61 & 1.56 & 0.68 & B & \nodata             & \nodata \\
HSC J2359$+$0208\D          & 359.8898 &    2.1399 & 0.44 & 0.430   &  8.67 & 2.88 & 0.35 & A & \textsc{Cam}RW      & $^2$ \\
\hline
\end{tabular}

%% file: tabs/Table4.tex
\begin{tabular}{L{2.4cm}C{2.1cm}C{1.1cm}C{1.1cm}}
\hline
Name &  Obs. Date & P.A.  & $\zs$  \\
     &  (UT)      & (deg)  &         \\
\hline
HSC J0224$-$0336                  & 13-07-2017 & 25                           & 1.514                  \\
HSC J0904$+$0125                  & 09-04-2017 & 5                             & 2.176                  \\
HSC J0907$+$0057                  & 29-01-2018 & 25                            & 1.916                  \\
HSC J1147$-$0013                  & 28-02-2018 & 102                            & 2.093                  \\
HSC J1156$-$0037                  & 07-04-2017 & 22                             & 1.907                  \\
HSC J1201$+$0126                  & 07-04-2017 & 7                            & 1.653                  \\
\multirow{2}{*}{HSC J1202$+$0039} & 06-04-2017 & \multirow{2}{*}{$-$55}       & \multirow{2}{*}{1.885} \\
	                                  & 01-03-2018 &        &                        \\
\multirow{2}{*}{HSC J2213$+$0048} & 10-06-2017 & \multirow{2}{*}{$-$18} & \multirow{2}{*}{\nodata} \\
                                  & 15-08-2017 &                                        &                        \\
HSC J2226$+$0041                  & 29-09-2017 & $-$50                 & 1.897                  \\
HSC J2233$-$0104                  & 07-08-2017 & 90                         & 0.902                  \\
\hline
\end{tabular}

%% file: tabs/TableA1.tex
\begin{tabular}{L{2cm}ccccC{2.2cm}}
\hline
    & \multicolumn{3}{c}{Grade} & \multirow{2}{*}{Total} & Previously known /\\
    & A & B & C &  & spectroscopically confirmed\\
\hline
SuGOHI-g           &   6 &  35 &  90 &  131 &  15 \\
\quad \textsc{Cam}&   1 &   5 &  32 &   38 &   3 \\
\quad F           &   2 &   4 &  13 &   19 &   7 \\
\quad R           &   0 &   5 &  11 &   16 &   3 \\
\quad W           &   0 &   8 &  14 &   22 &   4 \\
\quad Serendipity &   4 &  20 &  38 &   62 &   3 \\
\hline
\end{tabular}

%% file: tabs/TableA2.tex
\begin{tabular}{L{2.6cm}rrccccccC{1.8cm}C{1.7cm}}
\hline
Name & $\alpha$(J2000) & $\delta$(J2000) & $\zlp$ & $\zls$ & R$_{\rm arc}$ (arcsec) & Rank & $\rm \sigma_{Rank}$ & Grade & PC & References \\
\hline
HSC J0154$-$0039                &  28.6032 & $-$0.6610 & 0.18 & \nodata   &  1.31 & 1.89 & 0.74 & B & \nodata           & \nodata \\
HSC J0157$-$0500                &  29.3327 & $-$5.0109 & 0.28 & \nodata   &  1.32 & 2.22 & 0.42 & B & \textsc{Cam}RW    & \nodata \\
HSC J0208$-$0433\D              &  32.1339 & $-$4.5544 & 0.75 & \nodata   &  1.46 & 1.89 & 0.60 & B & F                 & $^8$ \\ 
HSC J0209$-$0244                &  32.4809 & $-$2.7451 & 0.56 & \nodata   &  1.10 & 3.00 & 0.00 & A & \nodata 			& \nodata \\
HSC J0214$-$0405\K              &  33.5467 & $-$4.0842 & 0.65 & 0.609     &  1.90 & 3.00 & 0.00 & A & F                 & $^{7,9}$ \\
HSC J0217$-$0513\K              &  34.4048 & $-$5.2249 & 0.64 & 0.646     &  1.71 & 3.00 & 0.00 & A & \textsc{Cam}F     & $^{3,9}$ \\
HSC J0218$-$0159                &  34.5994 & $-$1.9844 & 0.28 & \nodata   &  1.82 & 1.75 & 0.71 & B & \textsc{Cam}W     & \nodata \\
HSC J0218$-$0539                &  34.5983 & $-$5.6558 & 0.67 & 0.691     &  1.92 & 1.56 & 0.68 & B & F                 & \nodata \\
HSC J0228$-$0617                &  37.1722 & $-$6.2915 & 0.73 & \nodata   &  1.87 & 1.57 & 0.50 & B & \textsc{Cam}F     & \nodata \\
HSC J0233$-$0205                &  38.3446 & $-$2.0920 & 0.49 & \nodata   &  1.68 & 2.71 & 0.45 & A & \nodata 			& \nodata \\
HSC J0839$+$0210                & 129.8766 &    2.1733 & 0.67 & \nodata   &  1.91 & 1.56 & 0.68 & B & \nodata 			& \nodata \\
HSC J0904$-$0102\K              & 136.1239 & $-$1.0412 & 0.82 & 0.957     &  1.33 & 3.00 & 0.00 & A & \nodata 			& $^{20}$ \\
HSC J0913$+$0039$\dagger$\D     & 138.3797 &    0.6516 & 0.37 & 0.409     &  1.73 & 1.56 & 0.96 & B & \nodata 			& $^{4,5}$ \\
HSC J0923$+$0213                & 140.7907 &	2.2308 & 1.05 & \nodata   &  1.10 & 2.00 & 0.50 & B & \nodata 		    & \nodata \\ 
HSC J0925$+$0017                & 141.4375 &	0.2841 & 0.85 & \nodata   &  1.81 & 2.22 & 0.67 & B & \nodata 	        & \nodata \\
HSC J0959$+$0234                & 149.8789 &    2.5743 & 1.13 & \nodata   &  1.01 & 1.78 & 0.63 & B & \nodata 			& \nodata \\
HSC J1143$+$0040                & 175.8242 &    0.6760 & 0.36 & \nodata   &  1.71 & 1.67 & 0.47 & B & \nodata 			& \nodata \\
HSC J1201$-$0012\D              & 180.4162 & $-$0.2073 & 0.25 & \nodata   &  1.89 & 1.56 & 0.50 & B & R                 & $^4$ \\
HSC J1222$+$0205$\dagger$\D     & 185.6720 &    2.0995 & 0.24 & 0.229     &  1.85 & 1.75 & 0.71 & B & RW                & $^4$ \\
HSC J1234$-$0009\D              & 188.5910 & $-$0.1573 & 0.41 & \nodata   &  1.69 & 1.89 & 0.31 & B & \textsc{Cam}      & $^4$ \\
HSC J1344$-$0020                & 206.2318 & $-$0.3376 & 0.42 & \nodata   &  1.76 & 1.56 & 0.50 & B & \nodata 			& \nodata \\
HSC J1405$-$0028                & 211.2849 & $-$0.4751 & 0.56 & \nodata   &  1.78 & 1.63 & 0.52 & B & RW                & \nodata \\
HSC J1410$-$0109                & 212.7040 & $-$1.1630 & 0.66 & \nodata   &  1.07 & 2.44 & 0.73 & B & \nodata 			& \nodata \\
HSC J1413$-$0133                & 213.4980 & $-$1.5600 & 0.96 & \nodata   &  1.51 & 1.56 & 0.68 & B & \nodata 			& \nodata \\
HSC J1418$-$0003                & 214.7199 & $-$0.0660 & 0.55 & \nodata   &  1.46 & 1.89 & 0.60 & B & \nodata 			& \nodata \\
HSC J1420$+$0059                & 215.0560 &	0.9905 & 0.96 & \nodata   &  1.45 & 1.78 & 0.67 & B & \nodata 	        & \nodata \\
HSC J1421$+$0012                & 215.3635 &    0.2015 & 0.55 & \nodata   &  1.30 & 1.89 & 0.57 & B & \nodata 			& \nodata \\
HSC J1443$-$0007                & 220.9792 & $-$0.1252 & 0.89 & \nodata   &  1.29 & 2.11 & 0.31 & B & \nodata 			& \nodata \\
HSC J1501$+$4221                & 225.3007 &   42.3538 & 0.27 & \nodata   &  1.30 & 2.78 & 0.42 & A & \nodata 			& \nodata \\
HSC J1544$+$4427                & 236.2224 &   44.4637 & 0.66 & \nodata   &  0.94 & 1.67 & 0.47 & B & \nodata 			& \nodata \\
HSC J1608$+$4200                & 242.0651 &   42.0026 & 0.64 & 0.615     &  1.85 & 1.56 & 0.50 & B & W                 & \nodata \\
HSC J1613$+$5406                & 243.4006 &   54.1154 & 1.13 & 0.766     &  1.29 & 2.33 & 0.67 & B & \nodata 			& \nodata \\
HSC J1618$+$5449                & 244.6470 &   54.8230 & 0.81 & \nodata   &  1.60 & 1.56 & 0.68 & B & \textsc{Cam}      & \nodata \\
HSC J1640$+$4214                & 250.0049 &   42.2439 & 0.66 & \nodata   &  1.97 & 1.56 & 0.50 & B & \nodata 			& \nodata \\
HSC J2208$+$0446$\dagger$       & 332.0220 &    4.7676 & 0.27 & \nodata   &  1.60 & 1.56 & 0.50 & B & W                 & \nodata \\
HSC J2212$-$0018                & 333.0950 & $-$0.3032 & 0.91 & \nodata   &  1.31 & 2.00 & 0.50 & B & F                 & \nodata \\ 
HSC J2236$+$0240                & 339.2255 &    2.6785 & 0.38 & \nodata   &  1.14 & 1.67 & 0.82 & B & \nodata 			& \nodata \\
HSC J2310$+$0247\D              & 347.5196 &    2.7999 & 0.37 & 0.390     &  1.75 & 1.56 & 0.68 & B & RW                & $^6$ \\
HSC J2323$-$0030\D              & 350.9419 & $-$0.5105 & 0.91 & \nodata   &  1.53 & 2.33 & 0.67 & B & \nodata 			& $^3$ \\
HSC J2324$+$0127$\dagger$       & 351.0397 &    1.4579 & 0.19 & 0.190     &  1.94 & 2.44 & 0.83 & B & W                 & \nodata \\
HSC J2331$+$0000                & 352.7770 &	0.0035 & 0.69 & \nodata   &  1.10 & 2.00 & 0.71 & B & \nodata 		    & \nodata \\
\hline
\end{tabular}